\documentclass{aa}

\usepackage{graphicx}
\usepackage{txfonts}
\usepackage{lscape}
\usepackage{subcaption}
\usepackage{placeins}
\usepackage{siunitx}
\usepackage{natbib}
\bibpunct{(}{)}{;}{a}{}{,}
\newcommand{\comment}[1]{}
\newcommand{\RNum}[1]{\uppercase\expandafter{\romannumeral #1\relax}}

\begin{document}

   \title{Radial profiles of lensed $z\sim 1$ galaxies on sub-kiloparsec scales\thanks{The reduced maps are only available in electronic form at the CDS via anonymous ftp to cdsarc.u-strasbg.fr (130.79.128.5) or via http://cdsweb.u-strasbg.fr/cgi-bin/qcat?J/A+A/}}
   \authorrunning{D. Nagy et al.}

   \author{David Nagy\inst{1}\fnmsep\thanks{email: david.nagy@unige.ch}, Miroslava Dessauges-Zavadsky\inst{1}, Johan Richard\inst{2}, Daniel Schaerer\inst{1} \and Françoise Combes\inst{3} \and Matteo Messa\inst{1,4} \and John Chisholm\inst{5}
          }

   \institute{Observatoire de Genève, Université de Genève, Versoix, Switzerland
        \and
             Université Lyon, Université Lyon1, ENS de Lyon, CNRS, Centre de Recherche Astrophysique de Lyon UMR5574, Saint-Genis-Laval, France
        \and
            LERMA, Observatoire de Paris, PSL Research Université, CNRS, Sorbonne Université, UPMC, Paris, France
        \and
            Department of Astronomy, Stockholm University, AlbaNova University Centre, SE-106 91 Stockholm, Sweden
        \and
            Department of Astronomy, The University of Texas at Austin, 2515 Speedway, Stop C1400, Austin, TX 78712, USA
             }


 
  \abstract
   {We study the spatially resolved physical properties of the Cosmic Snake arc in MACS J1206.2–0847 and the arc in Abell 0521 (A521). These are two strongly lensed galaxies at redshifts $z=1.036$ and $z=1.044$. We used observations of the Hubble Space Telescope (HST) and the Atacama Large Millimeter/submillimeter Array (ALMA). The former gives access to the star formation rate (SFR) and stellar mass ($M_\star$), and the latter to the H$_2$ molecular gas mass ($M_{\mathrm{mol}}$). HST and ALMA observations have similar angular resolutions of $0.15\arcsec-0.2\arcsec$, which with the help of strong gravitational lensing enable us to reach spatial resolutions down to $\sim \SI{30}{pc}$ and $\sim \SI{100}{pc}$ in these two galaxies, respectively. These resolutions are close to the resolution of observations of nearby galaxies. We study the radial profiles of SFR, $M_\star$, and $M_{\mathrm{mol}}$ surface densities of these high-redshift galaxies and compare the corresponding exponential scale lengths with those of local galaxies. We find that the scale lengths in the Cosmic Snake are about $\SI{0.5}{kpc}-\SI{1.5}{kpc}$, and they are 3 to 10 times larger in A521. This is a significant difference knowing that the two galaxies have comparable integrated properties. These high-redshift scale lengths are nevertheless comparable to those of local galaxies, which cover a wide distribution. The particularity of our high-redshift radial profiles is the normalisation of the $M_{\mathrm{mol}}$ surface density profiles ($\Sigma M_{\mathrm{mol}}$), which are offset by up to a factor of 20 with respect to the profiles of $z=0$ counterparts. The SFR surface density profiles are also offset by the same factor as $\Sigma M_{\mathrm{mol}}$, as expected from the Kennicutt-Schmidt law.}


   \maketitle
%

\section{Introduction}

Observations of high-redshift galaxies give unique insights into the formation processes of young galaxies. The star formation rate (SFR), stellar mass ($M_\star$), H$_2$ molecular gas mass ($M_{\mathrm{mol}}$), dust mass ($M_{\mathrm{dust}}$), and metallicity are key observables for studying the physical properties of galaxies. At high redshift, we generally only have access to the galaxy-integrated physical properties. However, determining spatially resolved physical properties is needed to help us understand the galactic structures and evolutionary history. In nearby spatially resolved galaxies, the study of the radial profiles of the cited observables provide important information about the star formation activity inside the galaxy, which is responsible for their $M_\star$ and metallicity build-up (e.g. \citealt{regan_bima_2001,tremonti_origin_2004,leroy_star_2008,tamburro_geometrically_2008,rosales-ortega_new_2012,sanchez_mass-metallicity_2013,erroz-ferrer_muse_2019}).

The main objective of measuring the radial profiles of key observables in nearby galaxies is to constrain star formation models. Because we lack a strict theory of star formation, establishing empirical laws, such as the Kennicutt-Schmidt (KS; e.g. \citealt{kennicutt_star_1989}), main sequence (MS; e.g. \citealt{speagle_highly_2014}) and mass-metallicity (e.g. \citealt{tremonti_origin_2004,maiolino_amaze_2008}) relations, represents the first step in galaxy modelling. The radial profiles, and more generally, high-resolution studies of galaxies, enable us to test and calibrate these empirical star formation recipes. It is timely to determine whether the empirical star formation laws valid in nearby galaxies hold at higher redshifts.

Measuring the radial distribution of the key galactic observables at high redshift may prove challenging because spatial scales below $\sim \SI{1}{kpc}$ need to be resolved. At $z>1$, we can therefore currently only study radial profiles with the help of gravitational lensing, which enables us to reach sub-kiloparsec resolutions in strongly lensed galaxies.

Massive galaxies or massive galaxy clusters bend space-time and act as spatial telescopes. Taking advantage of them allows us to probe background galaxies at increased spatial resolutions and at amplified luminosity (e.g. \citealt{richard_locuss_2010,jones_resolved_2010,bayliss_probing_2014,livermore_resolved_2015,patricio_kinematics_2018,cava_nature_2018}). Using the lens model, constrained based on the mass distribution of the lens galaxy or cluster, the magnification map of the lens in the image plane at various redshifts can be predicted, and the image of the lensed galaxy in the source plane can be reconstructed. Because the surface brightness is conserved through lensing, surface densities of physical quantities can be compared without the need to correct for lensing.

We present in this work the spatially resolved study of the physical quantities related to the star formation process (SFR, $M_\star$, $M_{\mathrm{mol}}$, and $M_{\mathrm{dust}}$) of two strongly lensed galaxies, the Cosmic Snake at $z=1.036$, with $M_\star = (4.0\pm 0.5)\times 10^{10}\,\si{M_\odot}$ and $\mathrm{SFR} = 30\pm 10\,\si{M_\odot.yr^{-1}}$ \citep{cava_nature_2018}, and A521 at $z=1.044$ with $M_\star = (7.4\pm 1.2)\times 10^{10}\,\si{M_\odot}$ and $\mathrm{SFR} = 26\pm 5\,\si{M_\odot.yr^{-1}}$ (as derived in this work). A detailed analysis of the stellar clumps and molecular gas clouds of the Cosmic Snake has been performed by \citet{cava_nature_2018} using the Hubble Space Telescope (HST), and by \citet{dessauges-zavadsky_molecular_2019} using the Atacama Large Millimeter/submillimeter Array (ALMA), respectively. The study of stellar clumps and molecular gas clouds in A521 will be presented in two forthcoming papers (Messa et al., in prep. and Dessauges-Zavadsky et al., in prep.) The study of the molecular and ionised gas kinematics of both galaxies has been presented in \citet{girard_towards_2019} using ALMA data and observations from \citet{patricio_kinematics_2018} with the ESO Multi Unit Spectroscopic Explorer (MUSE), respectively. Moreover, \citet{patricio_resolved_2019} has studied the metallicity gradient in the Cosmic Snake using data obtained with the Spectrograph for INtegral Field Observations in the Near Infrared (SINFONI).

For both galaxies, we combine observations obtained with HST and ALMA and observations of the Cosmic Snake obtained with SINFONI and the Herschel space observatory in order to measure the radial profiles of their surface densities of SFR ($\Sigma \mathrm{SFR}$), $M_\star$ ($\Sigma M_\star$), $M_{\mathrm{mol}}$ ($\Sigma M_{\mathrm{mol}}$), and $M_{\mathrm{dust}}$ ($\Sigma M_{\mathrm{dust}}$), and the corresponding scale lengths (assuming an exponential profile). We compare these radial profiles with those of 52 nearby galaxies from studies by \citet{regan_bima_2001}, \citet{leroy_star_2008}, and \citet{tamburro_geometrically_2008}, as well as one individual galaxy at $z=1.45$ from \citet{ushio_internal_2021}, and two stacks obtained for the high-redshift galaxies analysed in \citet{calistro_rivera_resolving_2018} ($z\sim 2-3$) and in \citet{jafariyazani_spatially_2019} ($z\sim 0.1-0.42$).

The paper is structured as follows: in Sect. \ref{sec:observation} we present the studied galaxies and their gravitational lens modelling, observations, and data reduction. In Sect. \ref{sec:analysis} we present the measurements of $\Sigma \mathrm{SFR}$, $\Sigma M_\star$, $\Sigma M_{\mathrm{mol}}$, and $\Sigma M_{\mathrm{dust}}$ for the Cosmic Snake and A521 galaxies. In Sect.\ref{sec:discussion} we present the derived radial profiles and scale lengths, and we discuss our results. Finally, in Sect. \ref{sec:conclusion} we give our conclusions.

Throughout this paper, we adopt the $\Lambda$-CDM cosmology with $H_0 = \SI{70}{km.s^{-1}.Mpc^{-1}}$, $\Omega_{\mathrm{M}} = 0.3$, and $\Omega_{\Lambda} = 0.7$. We adopt a \citet{salpeter_luminosity_1955} initial mass function.


    \begin{figure}
       \centering
       \includegraphics[width=0.49\textwidth]{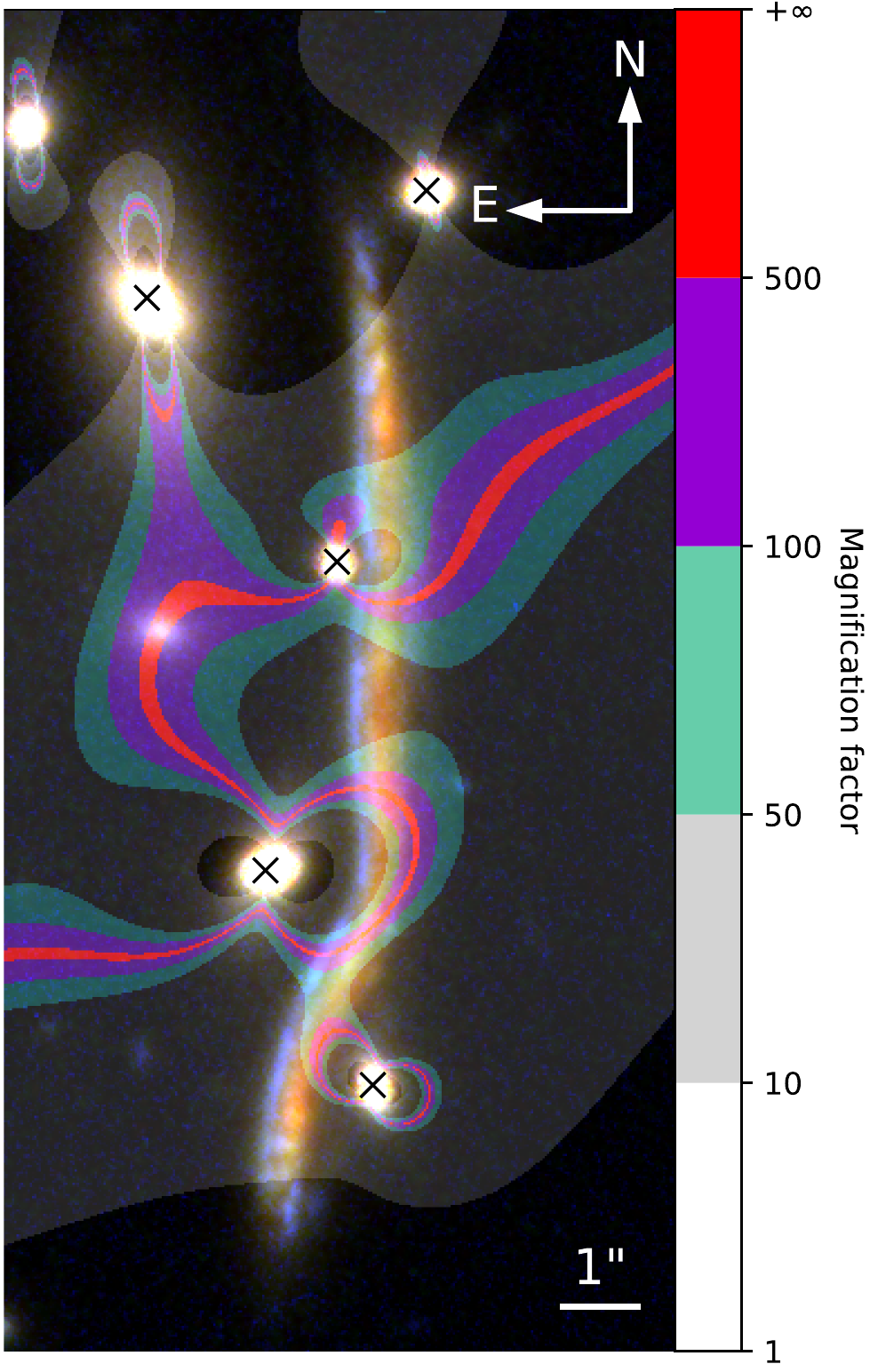}
          \caption{HST RGB-colour composite image of the Cosmic Snake (red filter: F160W, green: F105W, and blue: F606W). Magnifications are highlighted using coloured regions, from transparent for regions amplified 1 to 10 times, to red for regions magnified more than 500 times, as defined in the colour bar (on the right side). Galaxies marked with a cross are foreground lensing cluster member galaxies that contaminate the Cosmic Snake.
                  }
             \label{fig:snake_rgb}
       \end{figure}
      
     \begin{figure}
        \includegraphics[width=0.49\textwidth]{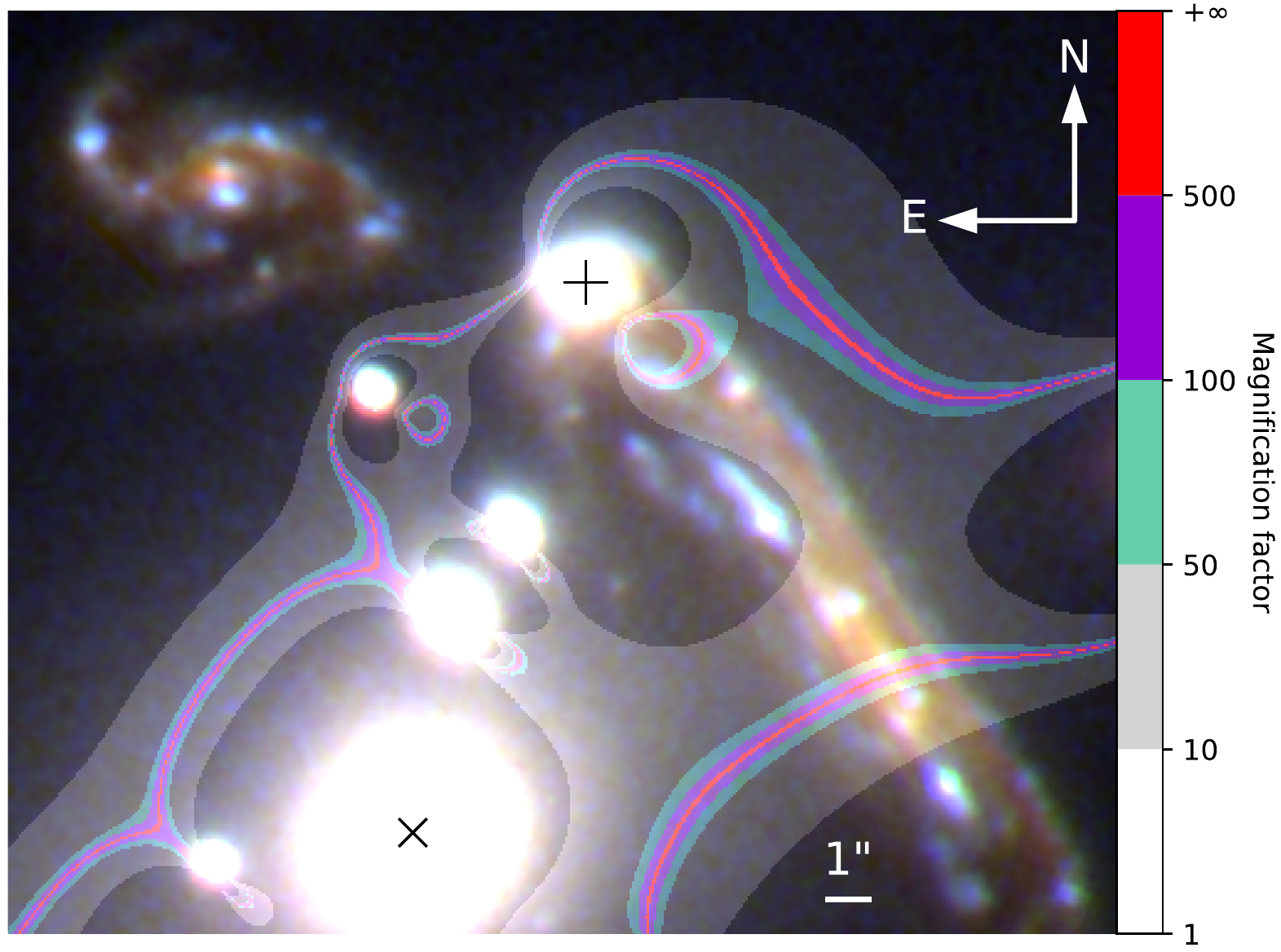}
      
        \caption{HST RGB-colour composite image of the A521 arc (red filter: F160W, green: F105W, and blue: F606W). Magnifications are highlighted using coloured regions, from transparent for regions amplified 1 to 10 times, to red for regions magnified more than 500 times, as defined in the colour-bar (on the right side). The galaxy marked with a plus is a foreground member of the lensing cluster that overlaps the A521 arc, and the galaxy marked with a cross is a foreground member that contaminates A521.} \label{fig:A521_rgb}
        \end{figure}

    \section{Observations and data reduction}
    \label{sec:observation}
    
    \subsection{Cosmic Snake and A521 galaxies}
    
    The Cosmic Snake arc in MACS J1206.2-0847 and the arc in Abell 0521 (which we refer to as A521) are two strongly lensed galaxies. We see several counter-images of them that are stretched and magnified from a few to hundreds of times (see Figs. \ref{fig:snake_rgb} and \ref{fig:A521_rgb}).
    
    The Cosmic Snake galaxy has four multiple images along an arc with magnifications between 10 and $>100$ (Fig. \ref{fig:snake_rgb}). All of them correspond to the northern part of the source galaxy. The northern and southern regions are images of slightly more than half of the galaxy, whereas the central region shows two highly magnified counter-images, both with fewer than 20\% of the galaxy. Slightly less than 50\% of the source galaxy is not visible in the arc. The Cosmic Snake galaxy has a fifth isolated counter-image to the north-east of the arc ($\sim \SI{15}{\arcsec}$ away from the northern tip of the arc), which is uniformly lensed by a magnification of 4.3 and is not shown in Fig. \ref{fig:snake_rgb}. With a total lensing-corrected $M_\star$ of $(4.0\pm 0.5)\times 10^{10}\,\si{M_\odot}$, a total lensing-corrected SFR of $30\pm 10 \,\si{M_\odot.yr^{-1}}$ , and a redshift of $z=1.03620$ \citep{cava_nature_2018}, the Cosmic Snake is representative of MS star-forming galaxies at $z\sim 1$ (e.g. \citealt{speagle_highly_2014}). The integrated $M_\star$ and SFR of the Cosmic Snake were measured on the isolated counter-image to the north-east, so that they also contain the $\sim 50\%$ of the mass and SFR that is absent in the arc.
    
    The A521 arc (Fig. \ref{fig:A521_rgb}) shows five counter-images of a source galaxy at $z=1.04356$ with magnifications between 3 and $>50$. The north-eastern counter-image is almost not stretched. This counter-image clearly shows that the A521 galaxy has the spiral structure typical of local late-type disky spiral galaxies. It is amplified by a factor growing from 2 on the easternmost part to 5-6 on the western spiral arm. Both spiral arms are visible in the second counter-image, which is the most elongated and magnified. The third counter-image in the south-west shows only a small part of the source galaxy, and an even smaller part of the source galaxy has two counter-images located beside (south) the cluster member galaxy marked with a plus in Fig. \ref{fig:A521_rgb}. A521 has an $M_\star$ of $(7.4\pm 1.2)\times 10^{10}\,\si{M_\odot}$ and an SFR of $26\pm 5\,\si{M_\odot.yr^{-1}}$. These quantities were derived from the spectral energy distribution (SED) fit on the HST photometry (see Sect. \ref{sec:HSTdataA521}) measured on the north-eastern counter-image. A521 is also representative of MS star-forming galaxies at $z\sim 1$.
    
    \subsection{HST observations}

    The HST images of MACS J1206.2–0847 are part of the Cluster Lensing and Supernova survey with Hubble (CLASH)\footnote{The data from CLASH are available at \url{https://archive.stsci.edu/prepds/clash/.}}. The Cosmic Snake was observed in 16 HST bands (4 WFC3/UVIS, 7 ACS, and 5 WFC3/IR). We used the $\SI{0.03}{^{\prime\prime}}$ version of the maps from the CLASH archive. The exposure time in each HST filter ranges between $\sim \SI{2000}{s}$ and $\sim \SI{5000}{s}$. The full CLASH dataset is described in \citet{postman_cluster_2012}.

    We took the A521 images observed in four HST bands (F390W with WFC3/UVIS, F105W and F160W with WFC3/IR, and F606W with WFPC2) from the HST archive (projects 11312 and 15435). The exposure time in each HST filter ranges between $\sim \SI{2000}{s}$ and $\sim \SI{5000}{s}$. Individual calibrated exposures were aligned and combined in a single image using the software Multidrizzle \citep{koekemoer_cosmos_2007}. The final images have spatial resolutions of $\SI{0.06}{^{\prime\prime}}$ and $\SI{0.1}{^{\prime\prime}}$ for WFC3 and WFPC2 observations, respectively. HST RGB-colour composite images including the magnification maps of the Cosmic Snake and A521 are displayed in Fig. \ref{fig:snake_rgb} and Fig. \ref{fig:A521_rgb}, respectively.

    \subsection{CO and dust continuum observations with ALMA}
    
    {The observations of the Cosmic Snake arc were acquired in Cycle 3 (project 2013.1.01330.S). They were performed in the extended C38-5 configuration with a maximum baseline of $\SI{1.6}{km}$ and 38 antennas of the 12 m array in band 6 with a total on-source integration time of $\SI{52.3}{min}$ \citep{dessauges-zavadsky_molecular_2019}. The targeted CO(4--3) emission line was observed at $\SI{226.44}{GHz}$, corresponding to a redshift of $z=1.036$. The observations of the isolated counter-image to the north-east of the Cosmic Snake arc and A521 were obtained in Cycle 4 (project 2016.1.00643.S). As described in \citet{girard_towards_2019}, the configuration we used was C40-6 with a maximum baseline of $\SI{3.1}{km}$ and 41 antennas of the 12 m array. The total on-source times in band 6 were  $\SI{51.8}{min}$ for the isolated counter-image of the Cosmic Snake and $\SI{89.0}{min}$ for A521. The tuned frequency for the isolated counter-image was the same as for the Cosmic Snake arc. For A521, the frequency was tuned at $\SI{225.66}{GHz}$, which corresponds to the frequency of the CO(4--3) line at the redshift of $z=1.043$. For all three observations, the spectral resolution was set to $\SI{7.8125}{MHz}$.}

\begin{figure*}

      \begin{subfigure}{0.333\textwidth}
        \includegraphics[width=\textwidth]{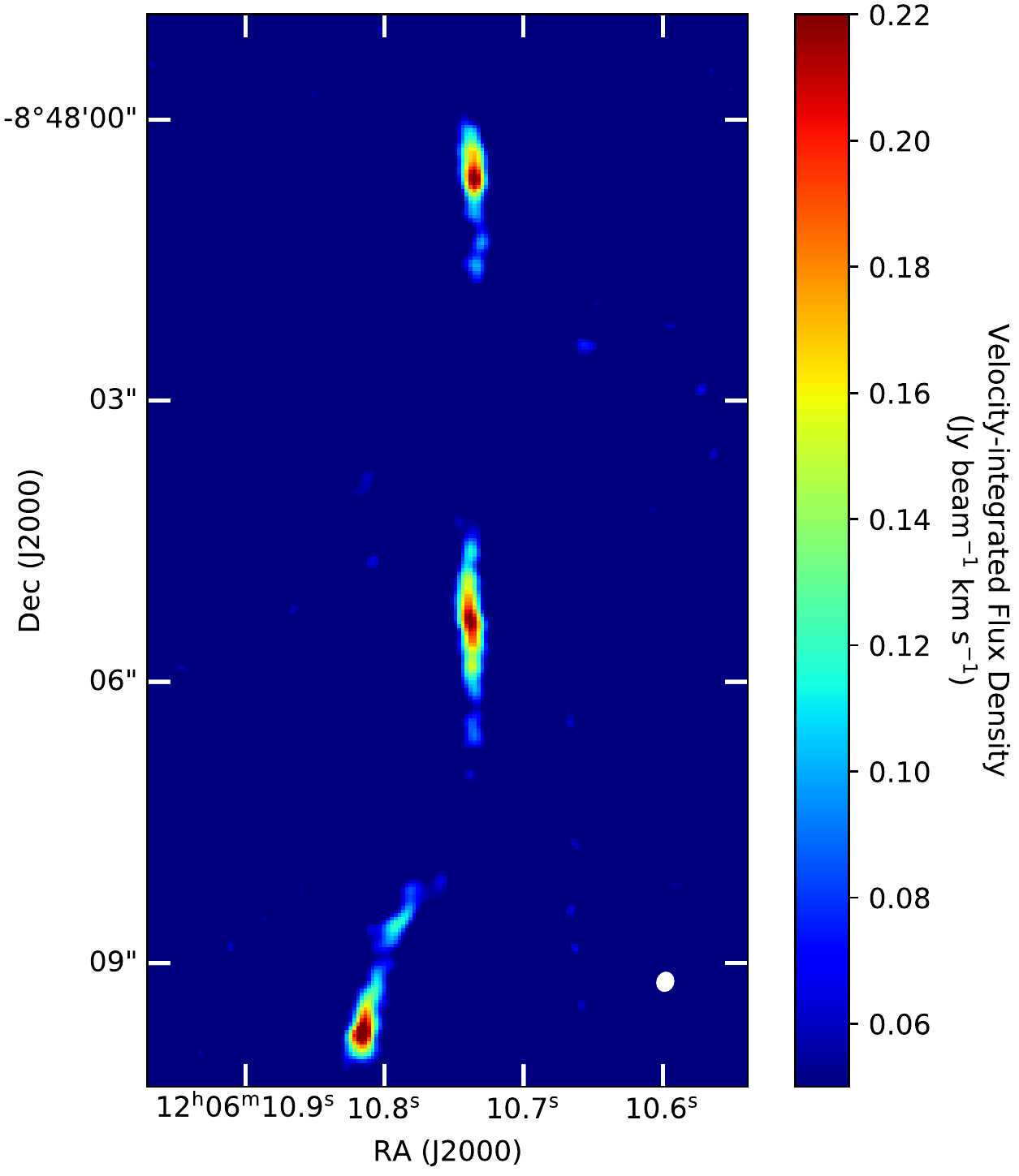}
      \end{subfigure}%
      \begin{subfigure}{0.333\textwidth}
        \includegraphics[width=\textwidth]{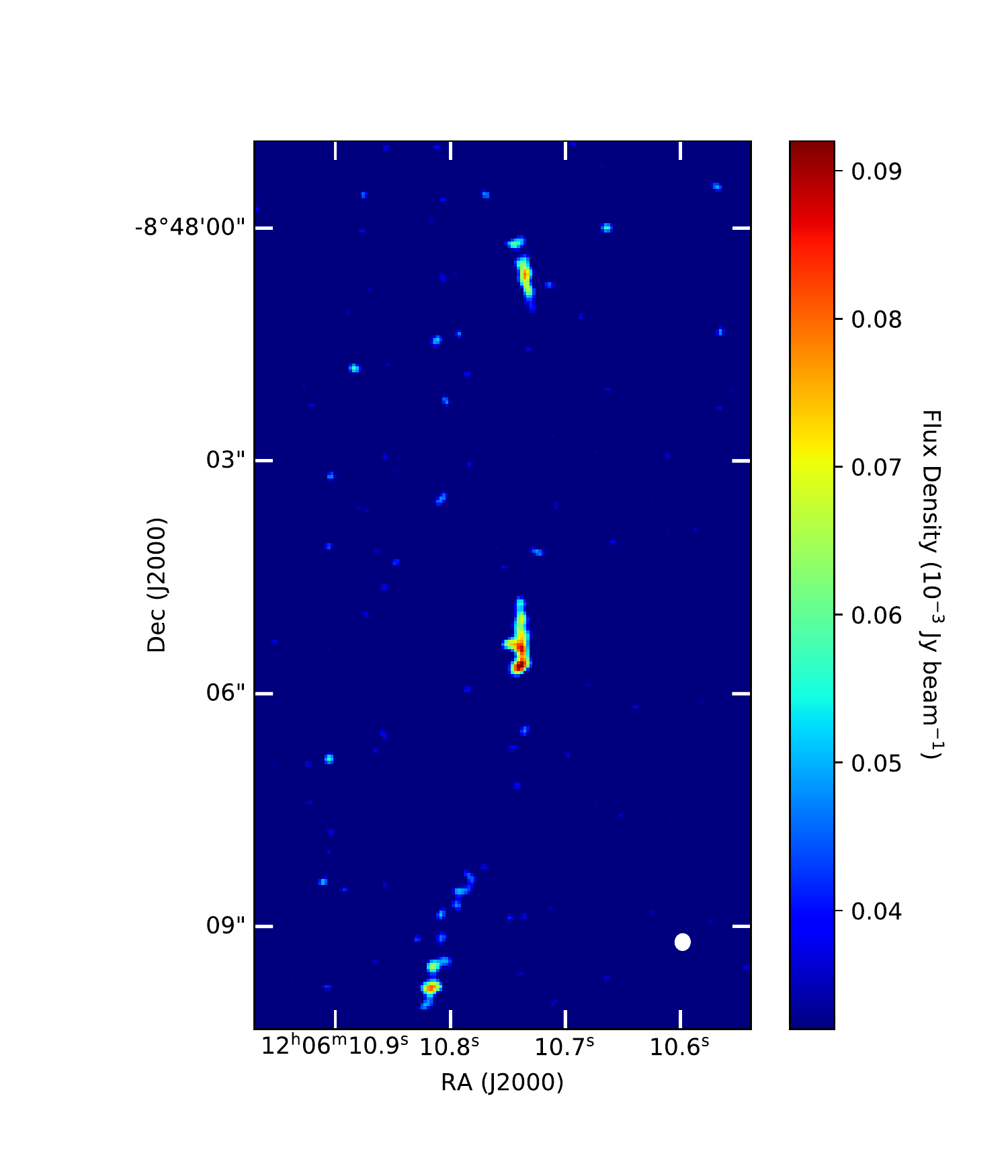}
      \end{subfigure}%
      \begin{subfigure}{0.333\textwidth}
        \includegraphics[width=\textwidth]{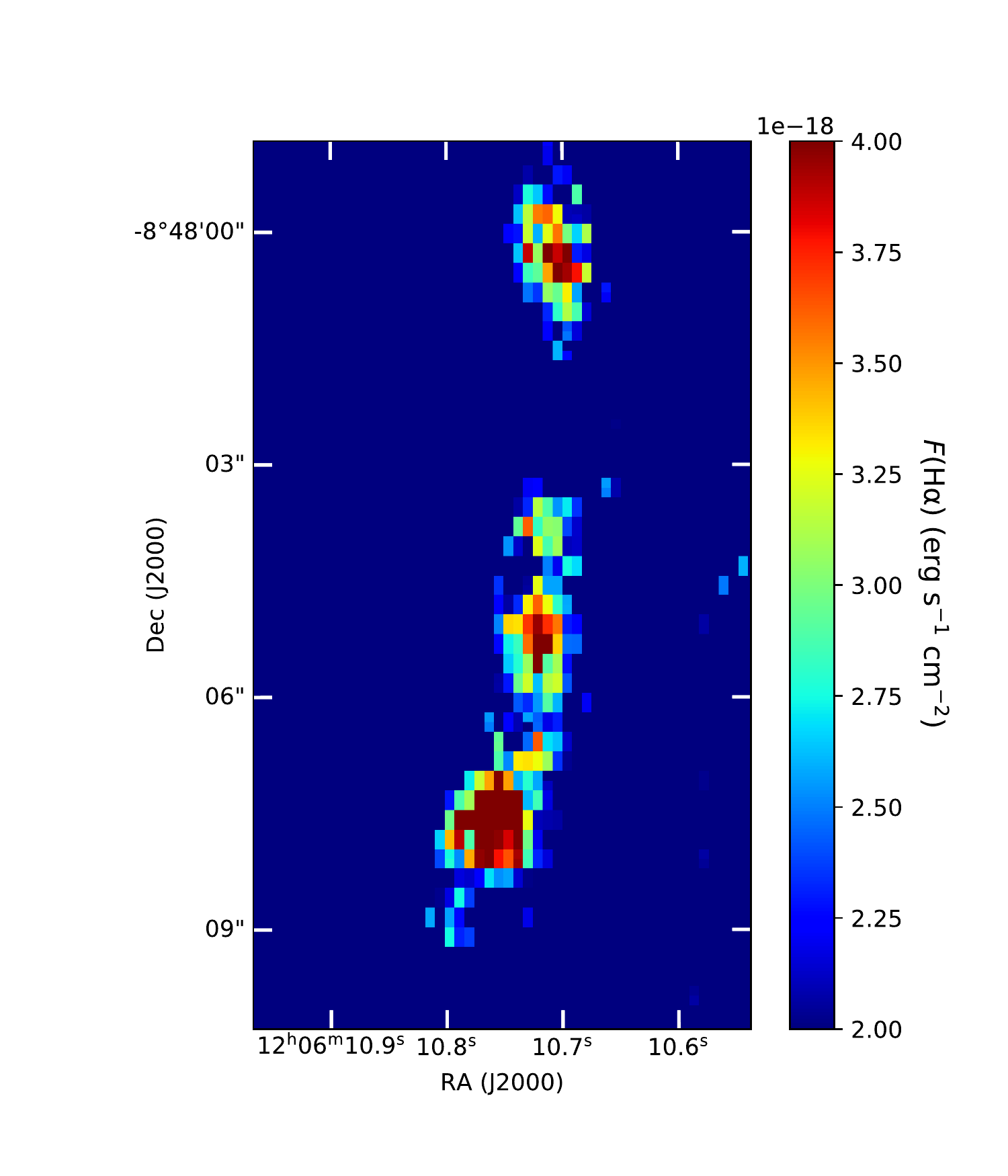}
      \end{subfigure}%
    
    \caption{ CO(4--3) moment-zero map (left) and rest-frame $\SI{650}{\mu m}$ continuum map (right) from ALMA. The colours indicate the values of the velocity-integrated flux density in $\si{Jy.beam^{-1}.km.s^{-1}}$ and the flux density in $\si{Jy.beam^{-1}}$. Right panel: $\mathrm{H\alpha}$ map of the Cosmic Snake from SINFONI. The colour bar indicates the $\mathrm{H_\alpha}$ flux ($F(\mathrm{H_\alpha})$) in $\si{erg.s^{-1}.cm^{-2}}$. The beam is displayed in white in the bottom right corners of the ALMA images.} \label{fig:snake}
    \end{figure*}

     For the data reduction, the standard automated reduction procedure from the pipeline of the Common Astronomy Software Application (CASA) package \citep{mcmullin_casa_2007} was used. Briggs weighting with a robust factor of 0.5 was applied to image the CO(4--3) emission. The cleaning was made interactively on all channels with the \emph{clean} routine in CASA until convergence, using a custom mask and ensuring that all the CO emission was included in all channels. A final synthesised beam size of $\SI{0.22}{\arcsec} \times \SI{0.18}{\arcsec}$ with a position angle of $85^\circ$ was obtained in the case of the Cosmic Snake arc, {$\SI{0.21}{\arcsec} \times \SI{0.18}{\arcsec}$ at $49^\circ$ for the Cosmic Snake isolated counter-image}, and $\SI{0.19}{\arcsec} \times \SI{0.16}{\arcsec}$ at $-74^\circ$ for A521. Primary beam correction was applied. The achieved RMS is $\SI{0.29}{mJy.beam^{-1}}$ per $\SI{7.8125}{MHz}$ channel for the Cosmic Snake arc, {$\SI{0.42}{mJy.beam^{-1}}$ per $\SI{7.8125}{MHz}$ channel for the isolated counter-image}, and $\SI{0.26}{mJy.beam^{-1}}$ per $\SI{7.8125}{MHz}$ channel for A521. The \emph{immoments} routine from CASA gave the CO(4--3) moment-zero maps by integrating the flux over the total velocity range showing the CO(4--3) emission. The resulting velocity-integrated CO intensity map of the Cosmic Snake arc is displayed in the middle panel of Fig. \ref{fig:snake}, and that of A521 is shown in Fig. \ref{fig:A521_alma}.

    We detect the $\SI{1.3}{mm}$ dust continuum in the Cosmic Snake arc, which corresponds to a rest-frame wavelength of $\SI{650}{\mu m}$. It was imaged over the calibrated visibilities of the four spectral windows and by excluding channels where the CO(4--3) emission was detected. The final synthesised beam size is of $\SI{0.20}{\arcsec} \times \SI{0.18}{\arcsec}$ at $86^\circ$ and RMS of $\SI{14}{\mu Jy.beam^{-1}}$. The $\SI{1.3}{mm}$ continuum map is displayed in the right panel of Fig. \ref{fig:snake}. No $\SI{1.3}{mm}$ continuum is detected in A521.
    
    \subsection{CO observations with PdBI}
    
    The Cosmic Snake was observed with the Plateau de Bure interferometer (PdBI) at $\SI{113.23}{GHz}$, the redshifted frequency of the CO(2--1) line. The observations were conducted in March 2014 using six antennas in the extended A-configuration for an on-source integration time of 7.6 hours. The data reduction was performed using the GILDAS software packages \emph{clic} and \emph{mapping}. The data were imaged with the \emph{clean} procedure (using the hogbom deconvolution algorithm) and natural weighting. The resulting synthesised beam after reduction has a size of $\SI{1.71}{\arcsec} \times \SI{0.62}{\arcsec}$ at the position angle of $17^\circ$, and the achieved RMS is $\SI{1.02}{m Jy.beam^{-1}}$ per $\SI{12}{MHz}$ channel.

    \subsection{$\mathrm{H\alpha}$ observations with SINFONI}

   The observations of the Cosmic Snake in $\mathrm{H\alpha}$ with SINFONI were performed in 2011, with an exposure time of 6h in total \citep{patricio_resolved_2019}. For the data reduction, the pipeline developed by MPE (SPRED; \citealt{abuter_sinfoni_2006,forster_schreiber_sins_2009}) was used in addition to custom codes for the correction of detector bad columns, cosmic ray removal, OH line suppression and sky subtraction \citep{davies_method_2007}, and flux calibration. The detailed steps of this procedure are explained in \citet{patricio_resolved_2019}. The $\mathrm{H\alpha}$ emission map of the Cosmic Snake is displayed in the left panel of Fig. \ref{fig:snake}. A521 was also observed with SINFONI \citep{patricio_resolved_2019}, but the galaxy is not bright enough in $\mathrm{H\alpha}$ to spatially study its emission.
    
    \subsection{Herschel data}
    For the Cosmic Snake we also considered the Herschel data from the Herschel Lensing Surveys \citep{egami_herschel_2010}. The Cosmic Snake is detected in all Herschel bands (PACS100, PACS160, SPIRE250, SPIRE350, and SPIRE500). Herschel photometry of the Cosmic Snake was carried out by \citet{cava_nature_2018}. A521 was also observed with Herschel, but we did not use Herschel data for A521.

     \begin{figure}
        \includegraphics[width=0.49\textwidth]{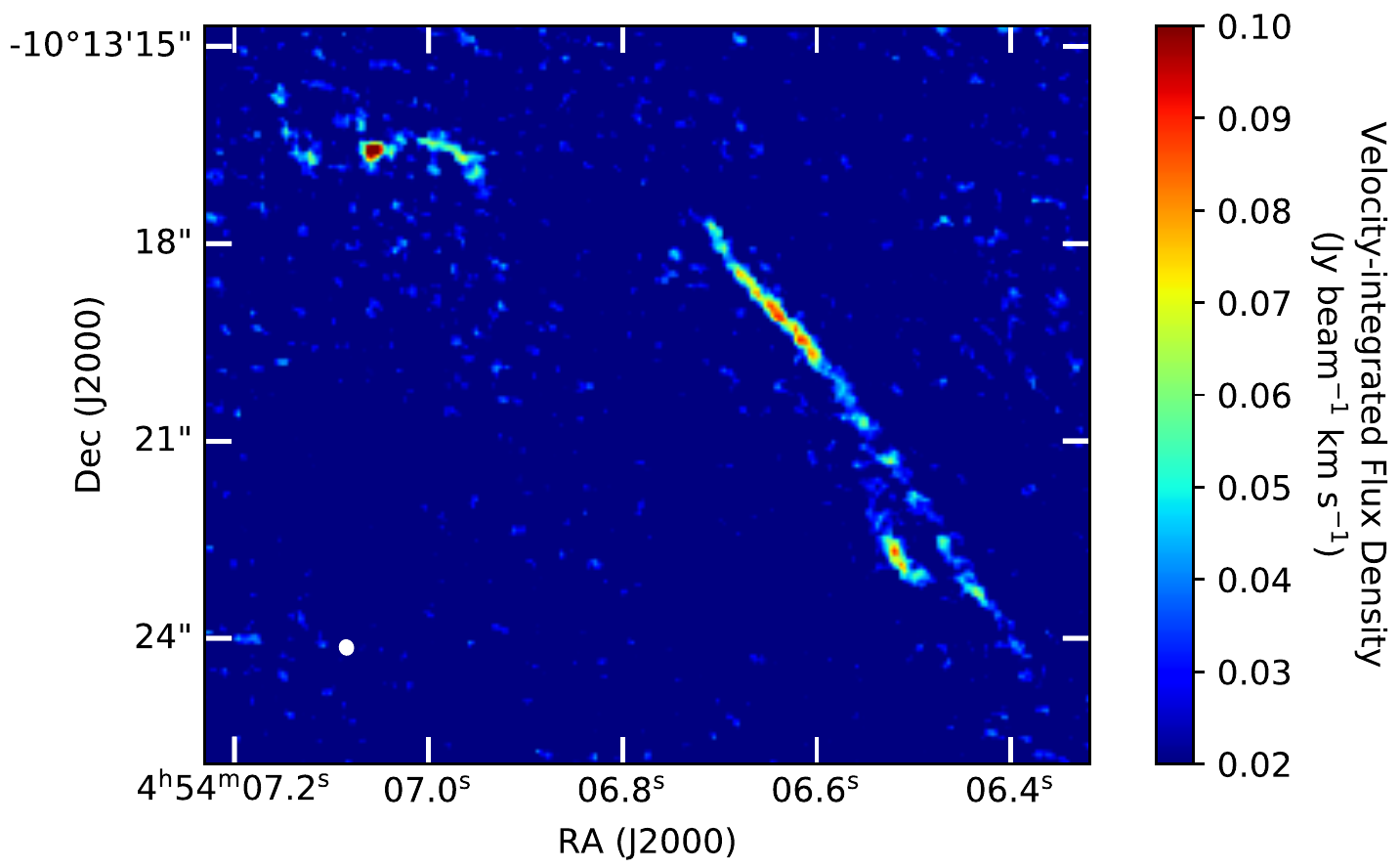}
      
    \caption{CO(4--3) moment-zero ALMA map of A521. The colours indicate the values of the velocity-integrated flux density in $\si{Jy.beam^{-1}.km.s^{-1}}$. The beam is displayed in white in the bottom left corner.} \label{fig:A521_alma}
    \end{figure}

    \subsection{Gravitational lens models}
 
    The gravitational lens models used for the Cosmic Snake and A521 are detailed in \citet{cava_nature_2018} and \citet{richard_locuss_2010}, respectively, and are constrained by multiple images observed in HST observations. The models were computed and optimised with Lenstool \citep{jullo_bayesian_2007}, resulting in magnification values with a precision between 80\% and 90\%. The root mean square (RMS) accuracy of the lens models for the positions in the image plane is 0.15\arcsec and 0.08\arcsec for the Cosmic Snake arc and A521 arc, respectively. These models are applied throughout the paper to recover the source properties of these lensed galaxies.

    \section{Determination of physical quantities}
    \label{sec:analysis}
    
    \subsection{Recovery of the galactocentric radius}
\begin{figure}
   \centering
   \includegraphics[width=0.49\textwidth]{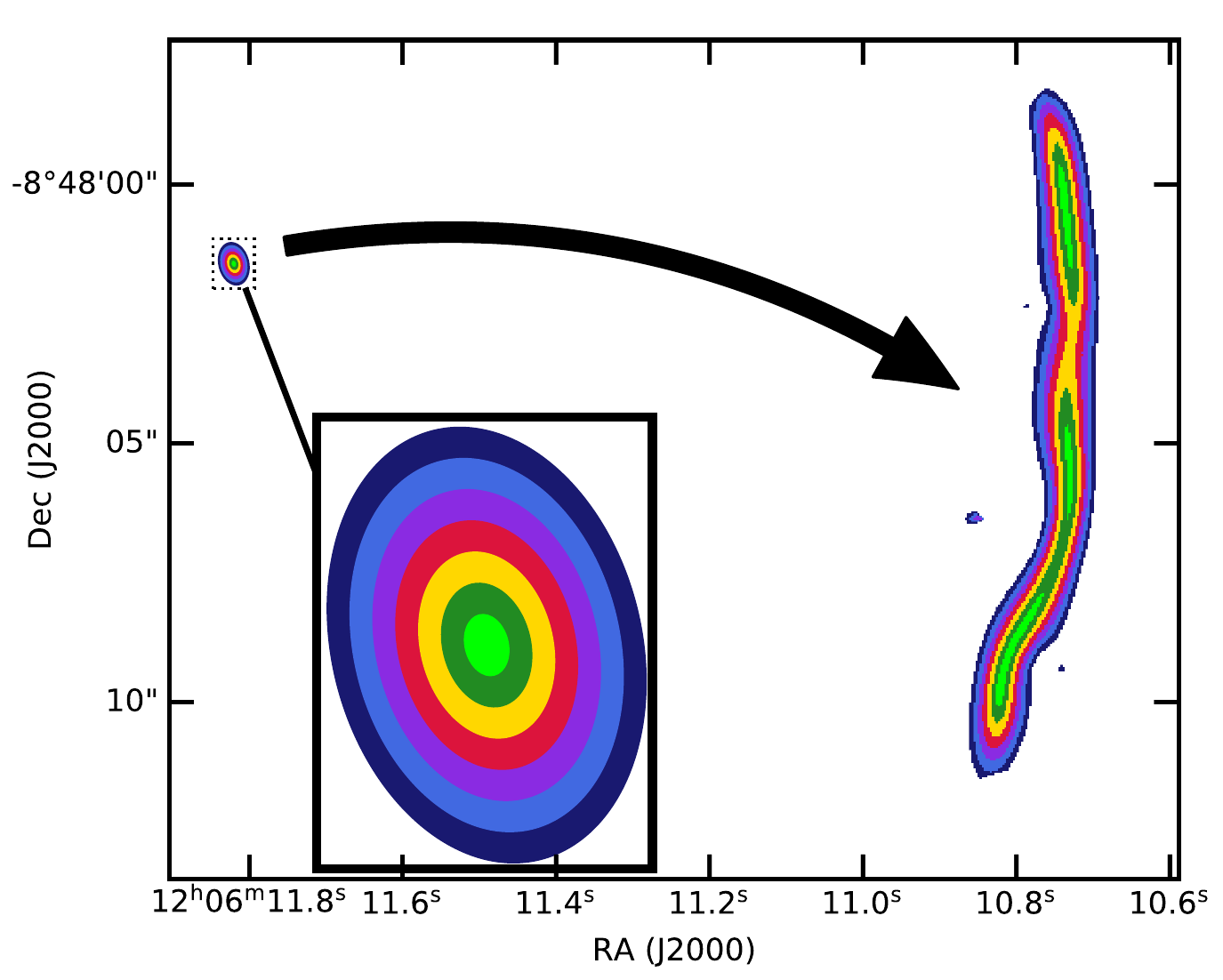}
      \caption{Colour-coded elliptical rings in steps of $\SI{0.5}{kpc}$ ranging from 0 to $\SI{3.5}{kpc}$ defined in the source plane image of the Cosmic Snake. The small ellipse is the size of the source plane reconstruction of the Cosmic Snake galaxy in the image plane. It is at the position where we would see it if it were not lensed. These elliptical rings are then projected over the Cosmic Snake arc in the image plane, on the right.}
         \label{fig:mask}
   \end{figure}

    In order to derive the galactocentric radii in the image plane, we proceeded as follows. We created a mask of concentric elliptical rings centred on the centre of the galaxy in the source plane (ellipses on the left of Fig. \ref{fig:mask}). Then, we projected this mask onto the image plane using the lens model of the galaxy, and obtained a map of galactocentric radii for the lensed image (arc on the right of Fig. \ref{fig:mask}). Fig. \ref{fig:mask} is an illustration of this process for the Cosmic Snake. Following the prescriptions of \citet{leroy_star_2008} for the local galaxies, we required that the width of the rings in the image plane was at least equal to half the size of the beam. We adjusted the width of the ellipses in the source plane accordingly. The number of ellipses was then determined by the maximum radius at which a given physical quantity of the galaxy was detected, and it therefore varies as a function of the data used (HST, ALMA, or SINFONI).

The surface of each ring ($S_{\rm ring}$) was computed following $S_{\mathrm{ring}} = N_{\mathrm{pix, ring}} \cdot p^2$, where $N_{\mathrm{pix, ring}}$ is the number of pixels in a given ring in the image plane and $p$ is the pixel size in physical units. {We assumed the surface brightness to be constant inside each ring, hence it is conserved through gravitational lensing. We can thus infer the surface density of various physical quantities without lensing corrections. To verify whether this hypothesis is correct, we derived for each quantity the lensing-corrected radial profile. To do this, we divided the brightness and the size of each pixel by its magnification obtained from the gravitational lens model when we integrated the flux inside an elliptical ring in the image plane. The resulting radial profiles deviate by less than $5\%$ for most data points and by never more than $10\%$ (only for a few data points) from the radial profiles obtained when a constant surface brightness was assumed. We are therefore confident that the assumption of a constant surface brightness in each ring is robust enough.} The following subsections describe how we inferred the various physical quantities for the Cosmic Snake and A521.

\subsection{Stellar mass and star formation rate surface densities}\label{sec:HSTdata}

\subsubsection{Cosmic Snake arc}\label{sec:HSTdataCS}

For the Cosmic Snake, we used the HST data from CLASH in 16 bands from the ultraviolet (UV) to the near-infrared \citep{cava_nature_2018} to determine its $\Sigma M_{\star}$ and $\Sigma \mathrm{SFR}$. The arc is especially in the redder bands ($\gtrsim \SI{700}{nm}$) contaminated by foreground cluster galaxy members. Before extracting the Cosmic Snake photometry, we subtracted the contamination from the brightest and closest cluster members to the arc (marked with a cross in Fig. \ref{fig:snake_rgb}). We computed the contamination by modelling the cluster members with a sum of two 2D Sérsic profiles each. For more details, see Appendix \ref{app:residuals}.

The Cosmic Snake galaxy is detected up to the galactocentric radius of $\SI{7}{kpc}$ in all HST bands. We defined a mask of 20 elliptical rings covering the range from $\SI{0}{kpc}$ to $\SI{7}{kpc}$ in the source plane with a step of $\SI{350}{pc}$, corresponding to our average HST spatial resolution. We measured the respective surface brightness in each ring and each band. The corresponding values are given in Table \ref{table:photometrysnake}.

We are interested in the radial profiles of $\Sigma \mathrm{SFR}$ and $\Sigma M_\star$, therefore we performed an SED fit on the photometry for each ring. From the SED fit, we also derived the attenuation ($A_{\mathrm{V}}$) that we used to correct the $\Sigma \mathrm{SFR}$ $\mathrm{H\alpha}$ radial profile for extinction (see Sect. \ref{sec:Halpha}), and the infrared luminosity ($L_{\mathrm{IR}}$), which we fine-tuned with the value measured in Sect. \ref{section:alma} to obtain a better constraint on the best SED fit.

We used the updated version of the HyperZ code for photometric redshift and spectral energy distribution \citep{schaerer_properties_2013} to fit the SED for each ring. We explored the fits performed with {16} different priors on the following parameters: the attenuation law, the age of the youngest stellar populations in the galaxy (Agemin), the minimum attenuation ($A_{\mathrm{V}}$ min), and the star formation history (SFH). The attenuation laws we used are the Milky Way law (MW) from \citet{allen_attempt_1976} and the Calzetti law from \citet{calzetti_dust_2000}. {We considered two SFHs: constant, and exponentially decreasing}. The priors are detailed in Table \ref{table:priors}, {where each model is tested with the two possible SFHs}. The combination of priors that yields the best match between the $L_{\mathrm{IR}}$ predicted by the SED fit and the measured lensed $L_{\mathrm{IR}}$ per ring\footnote{We assumed that all the absorbed (UV) flux is  reprocessed by dust locally, that is, in the same ring, meaning that we considered no leaking radiation outside the ring.} derived in Sect. \ref{section:alma} (see Table \ref{table:radprofLIR}) is model 4 from $\SI{0}{kpc}$ to $\SI{0.7}{kpc}$, model 3 from $\SI{0.7}{kpc}$ to $\SI{1.4}{kpc}$ kpc, and model 6 from $\SI{1.4}{kpc}$ to $\SI{7}{kpc}$, all adopting a MW attenuation law, $\mathrm{Agemin} = \SI{0.5}{Gyr}$, {and a constant SFH}. This indicates a strong gradient of attenuation from the central to the external regions of the Cosmic Snake galaxy. The HyperZ code yields errors on the derived $\Sigma \mathrm{SFR}$ and $\Sigma M_\star$, but our choice of priors is the main source of uncertainty here. We therefore assumed an error of $1 \sigma$ on the physical quantities derived from SED fits for each ring.

        \begin{table}
    \caption{Priors explored in the SED fits.}            
    \label{table:priors}
    \centering                        
    \begin{tabular}{c c c c}     
    \hline\hline                
        Model & Attenuation law & Agemin (Gyr) & $A_{\mathrm{V}}$ min \\
    \hline                        
        1 & MW & 0.1 & 2 \\
        2 & MW & 0.1 & 3 \\
        3 & MW & 0.5 & 2 \\
        4 & MW & 0.5 & 3 \\
        {5} & {MW} & {0.1} & {free}\\
        {6} & MW & 0.5 & free \\
        {7} & {Calzetti} & {0.1} & {free}\\
        {8} & Calzetti & 0.5 & free\\
    \hline                                
    \end{tabular}
    \tablefoot{{For each model above, we considered exponentially decreasing and constant SFHs.}}
    \end{table}

    \begin{table}
    \caption{Radial profile of the lensed $L_{\mathrm{IR}}$ for the Cosmic Snake.}             
    \label{table:radprofLIR}
    \centering                        
    \begin{tabular}{c c c}     
    \hline\hline                
        Radius (kpc) & $L_{\mathrm{IR}}$ ($\si{ .10^{10} .L_\odot}$)\\
    \hline                        
        0 & $10.2\pm 0.3$\\
        0.35 & $5.5\pm 0.3$\\
        0.7 & $2.5\pm 0.3$\\
        1.05 & $1.3\pm 0.3$\\
    \hline
    \end{tabular}
    \tablefoot{The values were computed assuming that the parameters $\nu_0$, $\beta,$ and $T_{\mathrm{dust}}$ of the MBB fitted over the total FIR SED of the Cosmic Snake are valid for the FIR SED of each ring.}
    \end{table}

\subsubsection{A521 arc}
\label{sec:HSTdataA521}

For A521, we used the HST data in four bands from the UV to the near-infrared to determine its $\Sigma M_{\star}$ and $\Sigma \mathrm{SFR}$. The arc, especially in the redder bands ($\gtrsim \SI{700}{nm}$), is contaminated by foreground cluster galaxy members. Five of them overlap the mask of A521. Three of them contaminate the mask only slighlty with their halos, which has a negligible impact on the photometry of A521. One of them (marked with a cross in Fig. \ref{fig:A521_rgb}) is particularly bright, therefore we modelled it with a Sérsic function and subtracted its flux from the image, as we did for the galaxies in front of the Cosmic Snake arc (see Appendix \ref{app:residuals}). The last member (marked with `+' on Fig. \ref{fig:A521_rgb}) overlaps the arc. Modelling this last cluster member would result in a large uncertainty. Therefore we instead cut out part of the mask at the location of the cluster galaxy, so that the contaminated pixels were left out from the flux integration. The cut was performed as follows: we traced a contour at $\mathrm{2\sigma}$ detection around the foreground galaxy and fitted an ellipse to it (using a least-squares algorithm). All pixels inside the fitted ellipse were excluded from the flux integration of the A521 arc. Because the contaminating cluster galaxy is detected in all bands, this process was applied for each band. To compute the uncertainty associated with the ellipse defining the pixels to be excluded from the flux integration over the A521 arc, we integrated the fluxes considering ellipses at $\mathrm{1\sigma}$ and $\mathrm{3\sigma}$ detection limits.

The A521 galaxy is detected up to the galactocentric distance of $\SI{20}{kpc}$ in the bands F606W, F105W, and F160W. However, in the HST band F390W, it is only detected up to $\SI{8}{kpc}$ . Therefore we used a mask covering the range from $\SI{0}{kpc}$ to $\SI{8}{kpc}$ of 20 elliptical rings with a step of $\SI{400}{pc}$ to match half of the average HST spatial resolution. We measured the respective surface brightness in each ring and each band. The corresponding values are given in Table \ref{table:photometryA521}.

We again performed SED fits for each ring to derive the radial profiles of $\Sigma \mathrm{SFR}$ and $\Sigma M_\star$. {We explored the different priors described in Table \ref{table:priors}. The SED fit with the best $\chi^2$ is obtained with model 6 and a constant SFH}. As we have no $L_{\mathrm{IR}}$ radial profile to constrain the SED fit for A521, we assumed that this choice of priors is valid at all radii. As for the Cosmic Snake, our choice of priors is the main source of uncertainty. We thus again assumed again an error of $\mathrm{1\sigma}$ on the physical quantities derived from SED fits for each ring.

To compute the total $M_\star$ and SFR of A521, we measured the total photometry on the north-eastern counter-image within the detection radius ($\SI{8}{kpc}$). {We then performed two SED fits using model 6 from Table. \ref{table:priors}}, the first assumed nebular emission, and the second assumed no emission. The two fits provide the upper and lower limits of our estimates, as the choice of priors is the main source of uncertainty. We obtain $M_\star = (7.4\pm 1.2)\times 10^{10}\,\si{M_\odot}$ and $\mathrm{SFR} = 26\pm 5\,\si{M_\odot.yr^{-1}}$. These values are lensing-corrected, assuming a constant magnification factor of $2.8$ on the counter-image. Our estimates are close to those of \citet{patricio_kinematics_2018}, who used the continuum of the MUSE spectrum of A521 to constrain $M_\star$ and SFR.

\subsection{Molecular gas and dust mass surface densities}
\label{section:alma}

The detected CO(4--3) line was used as the tracer of $M_{\mathrm{mol}}$ for both the Cosmic Snake and A521. To compute $M_{\mathrm{mol}}$ , we used the CO luminosity correction factor $r_{4,1}{=}L'_{\mathrm{CO(4 \mbox{--} 3)}}/L'_{\mathrm{CO(1 \mbox{--} 0)}}{=}0.33$, which we extrapolated from $r_{4,2}$ and $r_{2,1}$ measured in the Cosmic Snake \citep{dessauges-zavadsky_molecular_2019} and $z{\sim}1.5$ BzK galaxies \citep{daddi_co_2015}, respectively. We assumed a Milky Way CO-to-H$_2$ conversion factor $\alpha_{\mathrm{CO}}{=}\SI{4.36}{M_\odot(K. km. s^{-1}.pc^2)^{-1}}$ \citep{dessauges-zavadsky_molecular_2019}. We then have

   \begin{equation}
   \label{eq:mmol}
     M_{\mathrm{mol}} = \left(\frac{\alpha_{\mathrm{CO}}}{\si{M_\odot(K. km. s^{-1}.pc^2)^{-1}}}\right)\left(\frac{L'_{\mathrm{CO(4 \mbox{--} 3)}}/0.33}{\si{K. km. s^{-1}.pc^2}}\right)\si{M_\odot}
     ,\end{equation}
   
   \noindent where the CO(4--3) luminosity ($L'_{\mathrm{CO(4 \mbox{--} 3)}}$) is derived from the integrated flux of the CO(4--3) line \citep{solomon_molecular_1997}, 
   
       \begin{equation}
           L'_{\mathrm{CO(4 \mbox{--} 3)}} = 3.25 \times 10^{7} S_{\mathrm{CO(4 \mbox{--} 3)}}\Delta V\nu_{\mathrm{obs}}^{-2}D_{\mathrm{L}}^2(1+z)^{-3} (\si{K.km.s^{-1}.pc^2})
       ,\end{equation}
       
   \noindent with $S_{\mathrm{CO}}\Delta V$ the velocity-integrated flux in $\si{Jy.km.s^{-1}}$, $\nu_{\mathrm{obs}}$ the observed frequency in GHz, and $D_L$ the luminosity distance of the source in Mpc.
   
    We integrated the CO(4--3) flux from the moment-zero map in 6 rings with a width of $\SI{300}{pc}$ for the Cosmic Snake, and in 16 rings with a width of $\SI{400}{pc}$ for A521, corresponding to half of the average spatial resolution of the respective ALMA data. Because we detect the CO(4--3) line up to $\SI{1.7}{kpc}$ in the Cosmic Snake and {$\SI{6}{kpc}$} in A521, we only considered regions covering the corresponding galactocentric radii in the radial profiles of $\Sigma M_{\mathrm{mol}}$. The uncertainty ($\sigma_{\mathrm{ring}}$) on $\Sigma M_{\mathrm{mol}}$ in a given ring was computed following
   
   \begin{equation}
       \sigma_{\mathrm{ring}} = \frac{\sigma_{\mathrm{RMS}}}{\sqrt{N_{\mathrm{pix,ring}}/N_{\mathrm{pix,beam}}}}
   ,\end{equation}
   
    \noindent where $\sigma_{\mathrm{RMS}}$ is the RMS within the ring, $N_{\mathrm{pix,ring}}$ is the number of pixels in the ring, and $N_{\mathrm{pix,beam}}$ is the number of pixels in the beam.

    {To compute the total $M_{\mathrm{mol}}$ of the Cosmic Snake and A521 galaxies, we measured the CO(4--3) fluxes from the moment-zero maps of the isolated counter-image to the north-east of the Cosmic Snake arc within the detection radius of $\SI{1.7}{kpc}$, and the north-eastern counter-image of A521 within $\SI{6}{kpc}$. Using Eq. \ref{eq:mmol}, we obtain $M_{\mathrm{mol}} = (0.9\pm 0.3)\times 10^{10}\,\si{M_\odot}$ and $M_{\mathrm{mol}} = (1.1\pm 0.3)\times 10^{10}\,\si{M_\odot}$, respectively. These values are lensing-corrected, assuming constant magnification factors on the counter-images of the Cosmic Snake and A521 of $4.3$ and $2.8$, respectively.}

    From the rest-frame $\SI{650}{\mu m}$ dust continuum detected in the Cosmic Snake, we can determine $M_{\mathrm{dust}}$ from the observed flux density $S_{\mathrm{\nu}}$ following \citet{casey_far-infrared_2012},
    
    \begin{equation}
        M_{\mathrm{dust}} = \frac{S_{\mathrm{\nu}}D_{\mathrm{L}}^2}{\kappa B_{\mathrm{\nu}}(T_{\mathrm{dust}})}
    ,\end{equation}

   \noindent where the spectral radiance ($B_{\mathrm{\nu}}(T_{\mathrm{dust}})$) is computed from Planck's law at a dust temperature $T_{\mathrm{dust}}$ and at the observer-frame frequency ($\SI{230}{GHz}$). $\kappa=\kappa_{0}\left(\frac{\nu}{\nu_{0}}\right)^\beta$ is the dust mass absorption coefficient, where $\kappa_{0}$ is the dust opacity at $\nu_{0}$. $\beta$ is the spectral emissivity index\footnote{As mentioned in \citet{draine_dust_2007}, one should be cautious when computing dust masses at $\lambda < \SI{450}{\mu m}$ because Planck's law then becomes highly sensitive to $T_{\mathrm{dust}}$.}. $\nu_0$, $\beta$, and $T_{\mathrm{dust}}$ can be constrained by fitting a modified black-body (MBB) law over the complete FIR spectrum ($\sim 8-\SI{1000}{\mu m}$ in rest-frame) as in \citet{casey_far-infrared_2012},
    
        \begin{equation}
         S(\lambda) = N_{\mathrm{bb}}\frac{1-\exp\left(-\frac{\lambda_0}{\lambda}\right)^\beta}{\exp\left(\frac{hc}{k_BT_{\mathrm{dust}}\lambda}\right)}\left(\frac{c}{\lambda}\right)^3\label{eq:mbb}
        ,\end{equation}
        
    \noindent where $S(\lambda)$ is the flux density in wavelength space, $\lambda_0$ is the wavelength corresponding to the frequency $\nu_0$, and $N_{\mathrm{bb}}$ is the normalisation.
    
\begin{figure}
   \centering
   \includegraphics[width=0.49\textwidth]{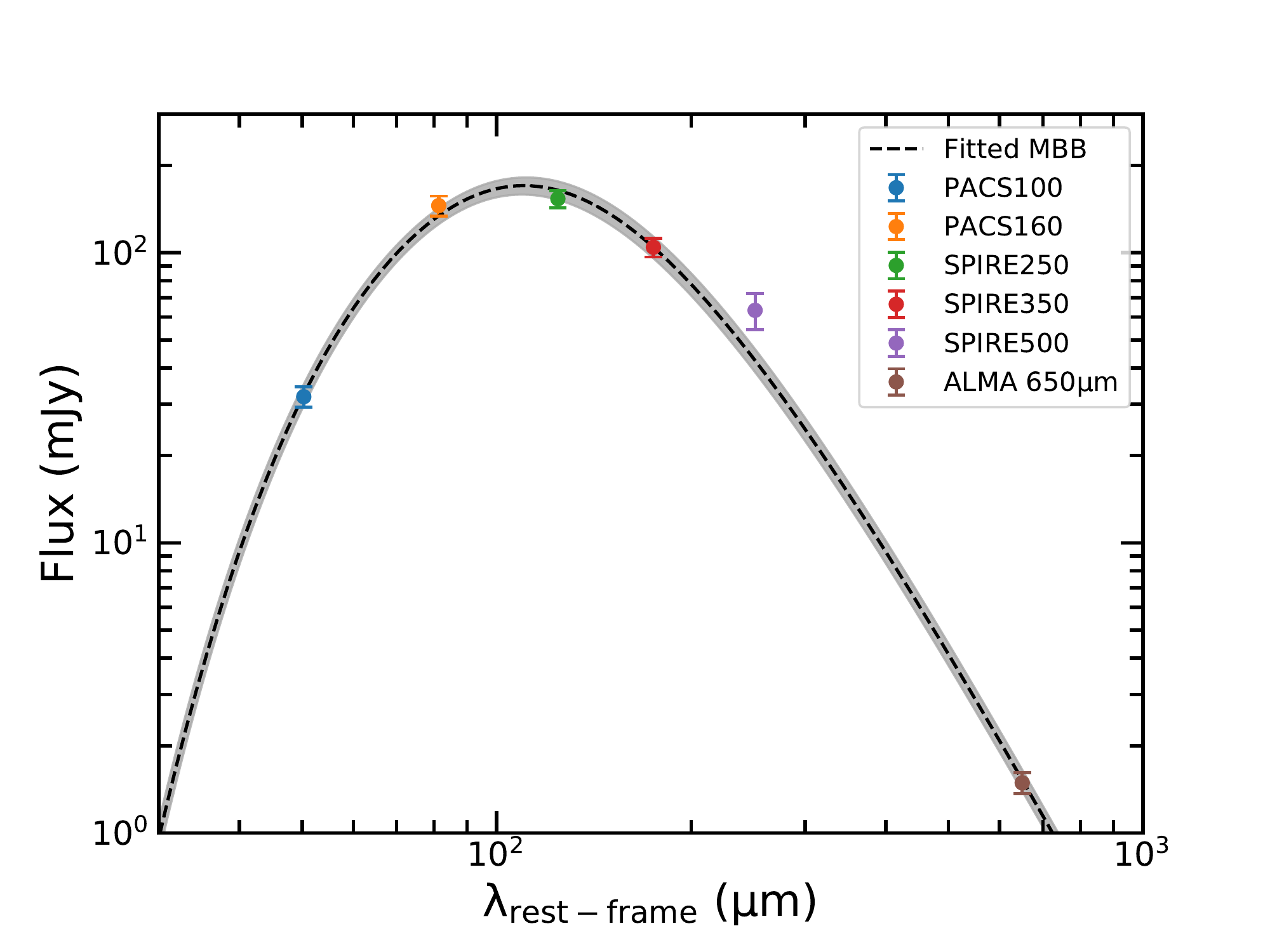}
      \caption{FIR SED spectrum of the Cosmic Snake. The dashed line is the best MBB fit on the Herschel data (PACS100, PACS160, SPIRE250, SPIRE350, and SPIRE500) and the ALMA rest-frame $\SI{650}{\mu m}$ measurement. The shaded area indicates the uncertainty on the fit.
              }
         \label{FigMBB}
   \end{figure}

    We performed a MBB fit over the integrated FIR SED of the Cosmic Snake, including the Herschel data and the ALMA rest-frame $\SI{650}{\mu m}$ measurement (Fig. \ref{FigMBB}). By integrating the fitted MBB over the FIR spectrum, we obtained the value of the total lensed infrared luminosity,
    
    \begin{equation}
    L_{\mathrm{IR}} = 4\pi D_L^2\int_{\lambda=\SI{8}{\mu m}}^{\SI{1000}{\mu m}} S_\nu d\nu = (885 \pm 65) \times 10^{10}\,\si{L_\odot}
    .\end{equation}

    \noindent The best fit yields values of $\beta = 2.2 \pm 0.2$, $\lambda_0 = c/\nu_0 = 162 \pm \SI{45}{\mu m}$, and $T_{\mathrm{dust}} = 38\pm \SI{3}{K}$. The errors reflect the uncertainties on the FIR photometric measurements. We assumed $\kappa_{0} = \SI{1.6}{m^2.kg^{-1}}$ as proposed in \citet{galliano_non-standard_2011} because the values of their corresponding $\beta$ and $\lambda_0$, 1.7 and $\SI{160}{\mu m}$, respectively, are close to those constrained by our fit.

    We then considered that the $\nu_0$, $\beta$, and $T_{\mathrm{dust}}$ parameters, derived for the total FIR SED of the Cosmic Snake, apply to the concentric radial rings, and we scaled the best-fitted FIR SED according to the ALMA rest-frame $\SI{650}{\mu m}$ dust continuum flux measurement in each ring in order to compute the radial profile of $\Sigma M_{\mathrm{dust}}$. We used a mask of {four rings} with a width of $\SI{300}{pc}$ ranging from $\SI{0}{kpc}$ to $\SI{1.2}{kpc}$ where we detect the rest-frame $\SI{650}{\mu m}$ continuum.
    
    To estimate the errors on $\Sigma M_{\mathrm{dust}}$, we computed values of $\Sigma M_{\mathrm{dust}}$ in each ring assuming several values of $\kappa$ calculated from different combinations of the parameters $\kappa_0$, $\lambda_0$ and $\beta$ as listed in \citet{ginolfi_infrared-luminous_2019}, and selected from the literature\footnote{The parameters are taken from \citet{bertoldi_dust_2003}, \citet{robson_submillimetre_2004}, \citet{beelen_350_2006}, \citet{weingartner_dust_2001}, \citet{bianchi_dust_2007}, \citet{jones_global_2017}, and \citet{galliano_non-standard_2011}.} assuming different dust composition models. We took the standard deviation of the resulting $\Sigma M_{\mathrm{dust}}$ in each ring as error, combined with the photometric error on the ALMA data.

    \subsection{$\mathrm{H\alpha}$ star formation rate surface density}
    \label{sec:Halpha}
    
    We used the SINFONI data mapping the $\mathrm{H\alpha}$ emission of the Cosmic Snake as the tracer of SFR. We considered the relation given in \citet{kennicutt_star_1998} to derive the SFR from the $\mathrm{H\alpha}$ luminosity ($L_{\mathrm{H\alpha}}$),

   \begin{equation}
     \mathrm{SFR} = \frac{L_{\mathrm{H\alpha}}}{\SI{1.26e41}{erg.s^{-1}}}\si{M_\odot.yr^{-1}}
   .\end{equation}
   
   We measured the $\mathrm{H\alpha}$ flux in five rings of $\SI{800}{pc}$ each, reflecting half of the average spatial resolution achieved in the SINFONI map. The rings cover the detection of $\mathrm{H\alpha}$ from $\SI{0}{kpc}$ to $\SI{4}{kpc}$. Using the radial profile of $A_{\mathrm{V}}$ derived in Sect. \ref{sec:HSTdataCS}, we corrected the $\mathrm{H\alpha}$ flux and then the $\mathrm{H\alpha}$ $\Sigma \mathrm{SFR}$ for extinction as in \citet{calzetti_dust_2000},

    \begin{equation}
        F_{\mathrm{obs}}(\lambda) = F_{\mathrm{int}}(\lambda)10^{-0.4A_\lambda} \label{eq:reddening}
    ,\end{equation}
    
    \noindent where $F_{\mathrm{obs}}$ and $F_{\mathrm{int}}$ are the observed and the intrinsic fluxes, and $A_{\mathrm{\lambda}}{=}\frac{k(\lambda) A_{\mathrm{V}}}{R_{\mathrm{V}}}$ with $k(\lambda{=}\SI{6563}{\angstrom})/R_{\mathrm{V}} {=}0.74$ for the MW attenuation law \citep{allen_attempt_1976}.

\section{Analysis and discussion}
\label{sec:discussion}
\subsection{Radial profiles and scale lengths of the Cosmic Snake and A521}
\label{sec:discussion-radprof}

\begin{figure*}
      \begin{subfigure}{0.5\textwidth}
   \includegraphics[width=\textwidth]{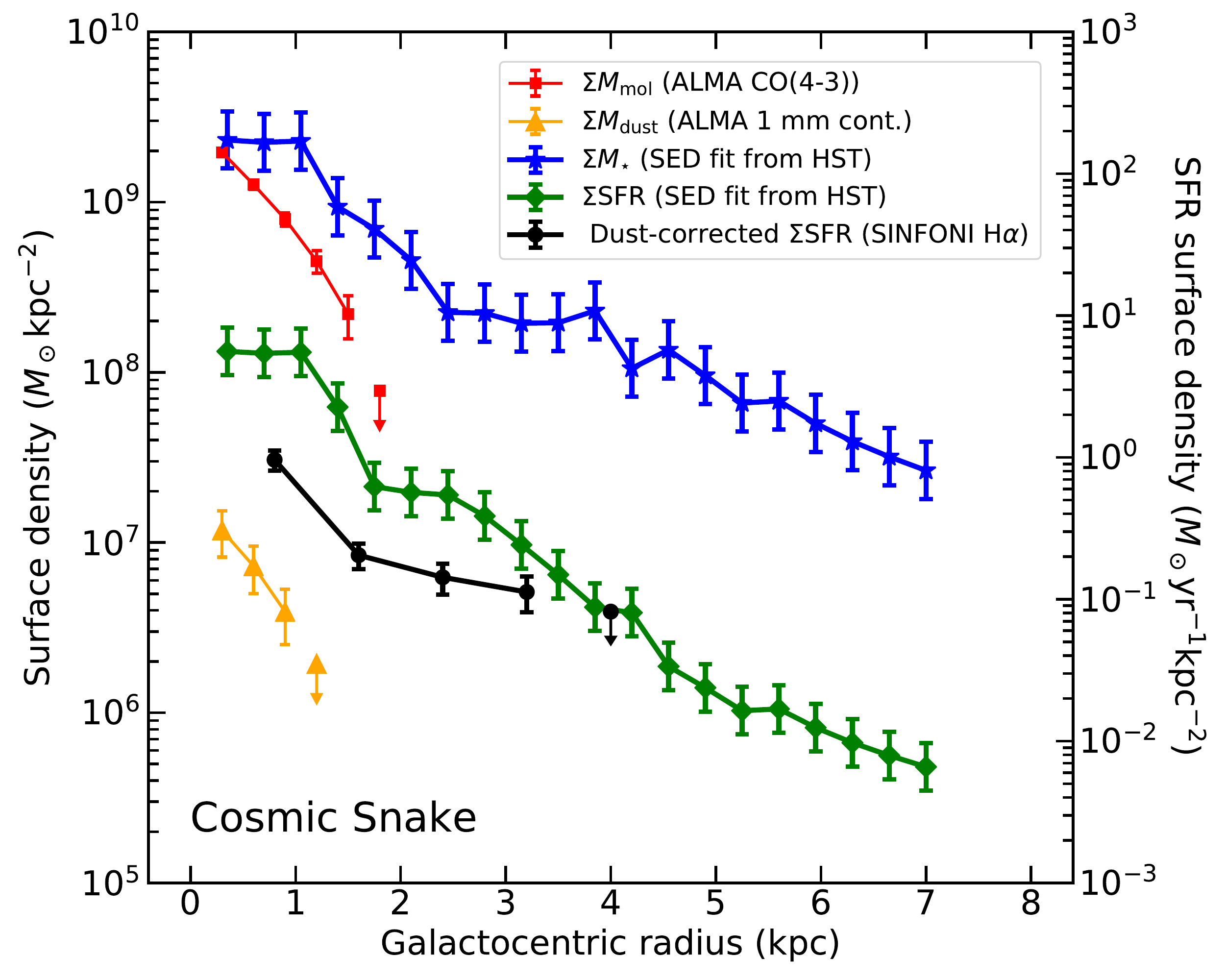}
      \end{subfigure}%
      \begin{subfigure}{0.5\textwidth}
   \includegraphics[width=\textwidth]{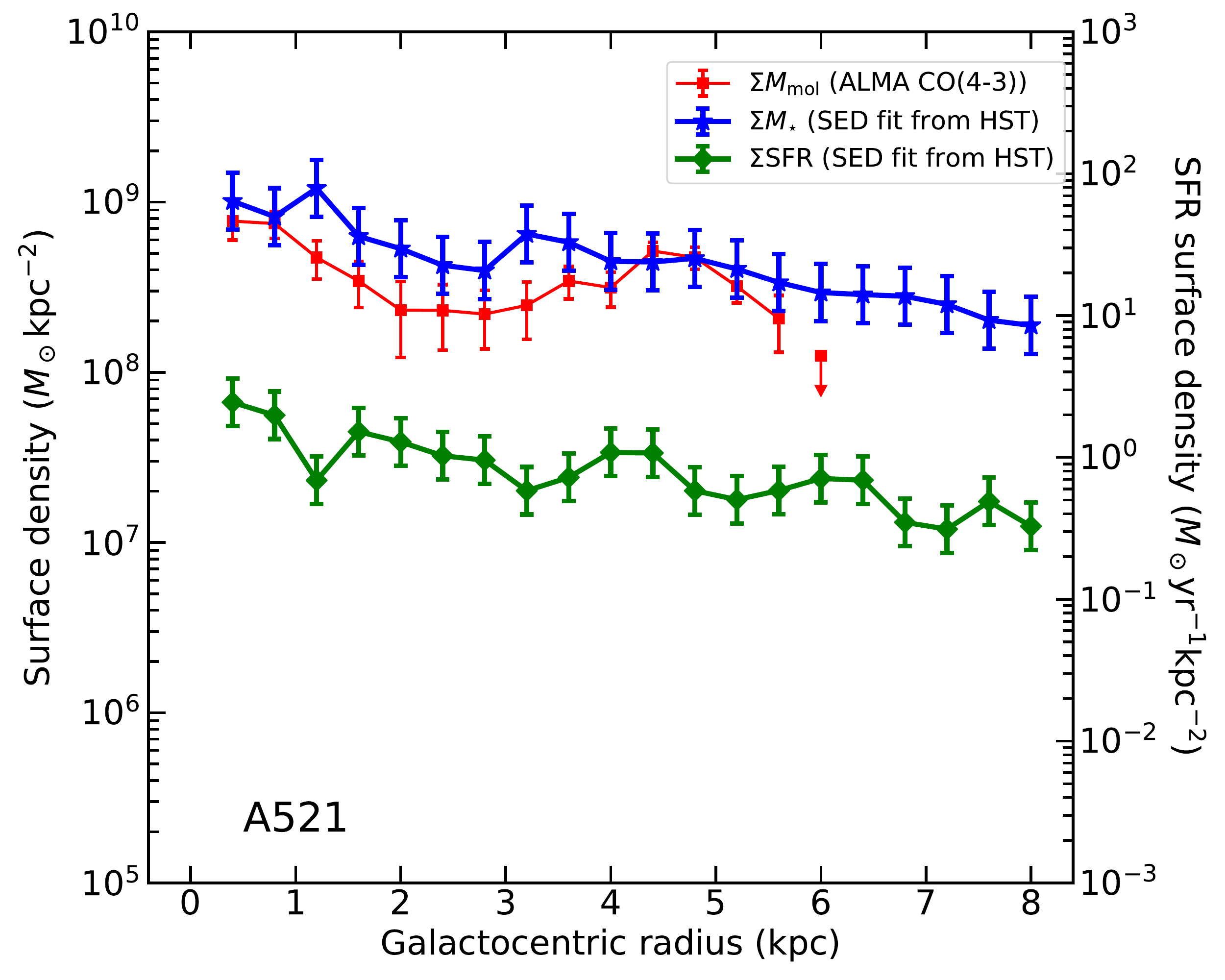}
      \end{subfigure}%
      \caption{Radial profiles of the Cosmic Snake (left panel) and A521 (right panel). The blue lines trace the $\Sigma M_\star$ profile, the red and orange lines trace the $\Sigma M_{\mathrm{mol}}$ and $\Sigma M_{\mathrm{dust}}$ profiles, and the green and black lines trace the $\Sigma \mathrm{SFR}$ radial profiles from the SED fits and the $\mathrm{H\alpha}$ emission, respectively. We consider as detection a signal above $\mathrm{4\sigma}$ for the ALMA data and above $\mathrm{2\sigma}$ for the other data. We indicate with an upper limit the datapoint below the corresponding detection threshold.}\label{fig:radialprofiles}
   \end{figure*}

We determined the radial profiles of $\Sigma M_\star$, $\Sigma \mathrm{SFR}$, $\Sigma M_{\mathrm{mol}}$, and $\Sigma M_{\mathrm{dust}}$. They are shown in Fig. \ref{fig:radialprofiles} for the Cosmic Snake in the left panel and for A521 in the right panel. These radial profiles can be characterised by a scale length $l$ if they are fitted by exponential functions. The scale length measures how far from the galactic centre a given quantity will be depleted on average. Uncertainties on scale lengths come from the fitting, and they include the error propagation from the flux measurements. The results are shown in Table~\ref{table:scalelength}.

Although the Cosmic Snake and A521 galaxies at $z\sim 1$ have comparable galaxy-integrated physical properties ($M_\star \sim (4-7.4) \times 10^{10} \mathrm{M_\odot}$, $\mathrm{SFR}\sim 26-\SI{30}{M_\odot.yr^{-1}}$, $M_{\mathrm{mol}} \sim (0.9-1.1) \times 10^{10}\,\si{M_\odot}$), their scale lengths are very different. The Cosmic Snake has much steeper profiles with scale lengths of about $0.5-\SI{1.5}{kpc}$, whereas A521 has longer scale lengths ($5-\SI{6}{kpc}$). We verified the robustness of this results with respect to the effects of spatial resolution and sensitivity.

    \begin{table}
    \caption{Scale length measurements of the Cosmic Snake ($l_{\mathrm{CS}}$) and A521 ($l_{\mathrm{A521}}$).}           
    \label{table:scalelength}
    \centering                        
    \begin{tabular}{c c c}     
    \hline\hline                
          & $l_{\mathrm{CS}}$ (kpc) & $l_{\mathrm{A521}}$ (kpc) \\
    \hline                        
        $\Sigma M_{\mathrm{mol}}$ & $0.55\pm 0.03$  & $6\pm 3$\\
        $\Sigma M_{\mathrm{dust}}$ & $0.54\pm 0.04$ & \\
        $\Sigma M_\star$ & $1.5\pm 0.1$ & $5.1\pm 0.5$\\
        $\Sigma \mathrm{SFR}$ ($\mathrm{H\alpha}$) & $1.0\pm 0.3$ & \\
        $\Sigma \mathrm{SFR}$ (SED) & $0.85\pm 0.06$ & $4.0\pm 0.4$\\
    \hline                                 
    \end{tabular}
    \end{table}

We convolved a given map at various resolutions and then used multiple elliptical ring spacings on the convolved maps and the original map to derive the corresponding scale length values. All values we found are within $\SI{25}{\%}$ of the scale length presented in this work. The spatial resolution therefore does not bias the measurement of the scale length because a difference of 25\% in our results would lead to the same conclusions given that the difference in scale lengths measured on our galaxies is much wider.

{We explored the possible effect of the sensitivity of the observations on the derived scale lengths. Because the sensitivity may affect the intrinsic size of a galaxy, we tested the variation of the measured scale length of a given radial profile detected up to different radii. We list in Table \ref{table:r_conv} the convergence radius ($r_{\mathrm{conv}}$), that is, the minimum radius up to which we need to detect the galaxy in order to find the measured scale lengths (within $1 \sigma$) given in Table \ref{table:scalelength}}. We compared it with the radius up to which we detected the galaxy ($r_{\mathrm{det}}$) for each physical quantity. All our $r_{\mathrm{conv}}$ are found to be below about half of the respective $r_{\mathrm{det}}$, with the exception of $\Sigma M_{\mathrm{mol}}$ in A521, which we discuss below. We conclude that the sensitivity of our observations is good enough to allow accurate scale length measurements.

    \begin{table}
    \caption{Convergence radii ($r_{\mathrm{conv}}$) and detection radii ($r_{\mathrm{det}}$) for the Cosmic Snake (CS) and A521}           
    \label{table:r_conv}
    \centering                        
    \begin{tabular}{c c c c}     
    \hline\hline                
          Galaxy & Quantity & $r_{\mathrm{conv}}$ (kpc) & $r_{\mathrm{det}}$ (kpc)\\
    \hline                        
        CS & $\Sigma M_{\mathrm{mol}}$ & 0.3 & {1.7} \\
        CS & $\Sigma M_{\mathrm{dust}}$ & 0.3 & {1.2} \\
        CS & $\Sigma M_\star$ & 3.5 & 7 \\
        CS & $\Sigma \mathrm{SFR}$ (SED) & 1.4 & 7 \\ 
        CS & $\Sigma \mathrm{SFR}$ ($\mathrm{H\alpha}$) & 2.4 & 4 \\                       
        A521 & $\Sigma M_{\mathrm{mol}}$ & 5.6 & 6 \\
        A521 & $\Sigma M_\star$ & 3.2 & 8 \\
        A521 & $\Sigma \mathrm{SFR}$ (SED) & 3.6 & 8 \\
    \hline                                 
    \end{tabular}
    \end{table}

In A521, we measured a scale length for $\Sigma M_{\mathrm{mol}}$ of $6\pm \SI{3}{kpc}$ by fitting an exponential on the entire set of points within the detection radius which, as we mentioned above, is much larger than in the Cosmic Snake (see Table \ref{table:scalelength}). However, in A521, the $\Sigma M_{\mathrm{mol}}$ radial profile does not follow a regular exponential, but shows a bump at $4-\SI{5}{kpc,}$ resulting in a high scale length value with a large uncertainty. This bump is due to the western spiral arm, which appears to be very bright in the CO(4--3) moment-zero map in the north-eastern counter-image and in the western counter-images (see Fig. \ref{fig:A521_alma}). Another way of measuring the scale length is to fit an exponential only on the points before the bump at radii from $\SI{0}{kpc}$ to $\SI{2.1}{kpc}$. In this way, we do not measure the global $\mathrm{H_2}$ depletion in the galaxy, but the internal scale decrease from the centre out to the spiral arm, for which we find $l_{\mathrm{mol}} = 1.3\pm \SI{0.2}{kpc}$. Even with this calculation, $l_{\mathrm{mol}}$ remains more than twice larger than in the Cosmic Snake.

In the Cosmic Snake, we compared the scale length of the CO(4--3) emission detected with ALMA with the scale length of the PdBI CO(2--1) emission to verify whether different CO transitions still show equally short scale lengths. To match the resolution of the CO(2--1) moment-zero map, the width of the rings should be about $\SI{2}{kpc}$. However, the CO(2--1) is detected up to $\sim\SI{2.5}{kpc}$. In order to keep two datapoints, we therefore adopted rings spaced by $\SI{1.2}{kpc}$, about half of the required width, although one should be aware that a smaller ring spacing might increase the dependence on noise and irregularities in the galaxy. We find a significant difference between the CO(4--3) and CO(2--1) scale lengths, as we measure a scale length for CO(2--1) of $\SI{4.3}{kpc}$. However, this result is very uncertain because we have only two datapoints for CO(2--1). We can nevertheless note that we detect the CO(2--1) emission farther out in galactocentric distance than the CO(4--3) emission (detected up to $\SI{1.7}{kpc}$). {This is in general expected because the CO(4--3) line emission, which traces denser gas, is confined to regions closer to the galactic centre.} Dust is also detected where dense molecular gas is observed, as the $\Sigma M_{\mathrm{dust}}$ profile follows the CO(4--3) $\Sigma M_{\mathrm{mol}}$ profile. Finally, the $\Sigma \mathrm{SFR}$ radial profiles derived from SED fitting and $\mathrm{H\alpha}$ line emission have the same scale lengths within error bars.{ This supports our conviction that the measurement of $l_\mathrm{SFR}$ for the Cosmic Snake is a robust result.}

\subsection{Morphological parameters of the Cosmic Snake and A521}
\label{sec:morphology}

The difference in the scale lengths of SFR and $M_{\mathrm{mol}}$ between the Cosmic Snake and A521 is expected because the shape of the radial profiles of these quantities can differ much between two galaxies, depending on their evolutionary stage. The molecular gas is accreted, then concentrates towards the centre because of torques caused by asymmetries, such as the bar or the spiral arms, on timescales of a few $10^8-10^9\,\si{yrs}$. This gas is consumed to form stars, therefore we expect various distributions of SFR and $M_{\mathrm{mol}}$ during the lifespan of a galaxy, with concentrations in the centre, in the spiral arms, or in Lindblad resonance rings \citep{lindblad_circulation_1964}. It has been suggested that as the molecular gas is compacted in the centre, the SFR is quenched, which then proceeds inside-out (e.g. \citealt{gonzalez_delgado_califa_2015,tacchella_evidence_2015,jafariyazani_spatially_2019}). Usually, ultra-luminous galaxies tend to have concentrated gas, whereas MS galaxies may show gas and SFR in the spiral arms or in the rings.

On the other hand, the $M_\star$ radial distribution in galaxies is much more constant. We do not expect large differences because the stars accumulate during the lifespan of a galaxy, either in the bulge and the disk, or only in the disk. We investigated the scaling relation between $\mu_\star^{\mathrm{HLR}}$ and $M_\star$ for the Cosmic Snake and A521 as in \citet{kauffmann_dependence_2003} and \citet{gonzalez_delgado_califa_2015}, where $\mu_\star^{\mathrm{HLR}}$ is the $\Sigma M_\star$ inside the half-light radius (HLR). To compute the HLR, we used the HST F160W filter map, corresponding to the rest-frame $z$ band as in \citet{kauffmann_dependence_2003}. We then measured the HLR directly from the radial profile of the flux computed on the isolated counter-image to the north-east for the Cosmic Snake and on the eastern counter-image for A521, {and corrected for lensing}. We obtain $3.5\pm 0.1\,\si{kpc}$ and $5.6\pm 0.2\,\si{kpc}$, respectively. The derived HLR of the Cosmic Snake and A521 are typical of galaxies at $z\sim 1$ with comparable $M_\star$ (see e.g. \citealt{van_der_wel_3d-hstcandels_2014}). We then used the radial profile of $\Sigma M_\star$ to compute $\mu_\star^{\mathrm{HLR}}$, and we find $(5.2\pm 0.5) \times 10^{8}\,\si{M_\odot.kpc^{-2}}$ for the Cosmic Snake and $(4.9\pm 0.2) \times 10^{8}\,\si{M_\odot.kpc^{-2}}$ for A521. Both the Cosmic Snake and A521 satisfy the relation between $\mu_\star^{\mathrm{HLR}}$ and $M_\star$ defined empirically by the distribution of local galaxies \citep{kauffmann_dependence_2003,gonzalez_delgado_califa_2015}. This suggests that even for high-redshift MS galaxies, the radial profile of $\Sigma M_\star$ mostly depends on their total $M_\star$. Furthermore, the $\mu_\star^{\mathrm{HLR}}$--$M_\star$ relation has been shown to secondarily depend on the morphological type of galaxies (e.g. \citealt{shimasaku_statistical_2001,gonzalez_delgado_califa_2015}). {According to Fig.~5 of \citet{gonzalez_delgado_califa_2015}, the Cosmic Snake and A521 appear to be of comparable morphological type (Sa-Sb), with the Cosmic Snake being tentatively of earlier type (Sa).}

\subsection{Comparison of radial profiles with local galaxies and other high-z objects}
\label{sec:comparison}

\begin{figure*}
      \begin{subfigure}{0.333\textwidth}{}
        \includegraphics[width=\textwidth]{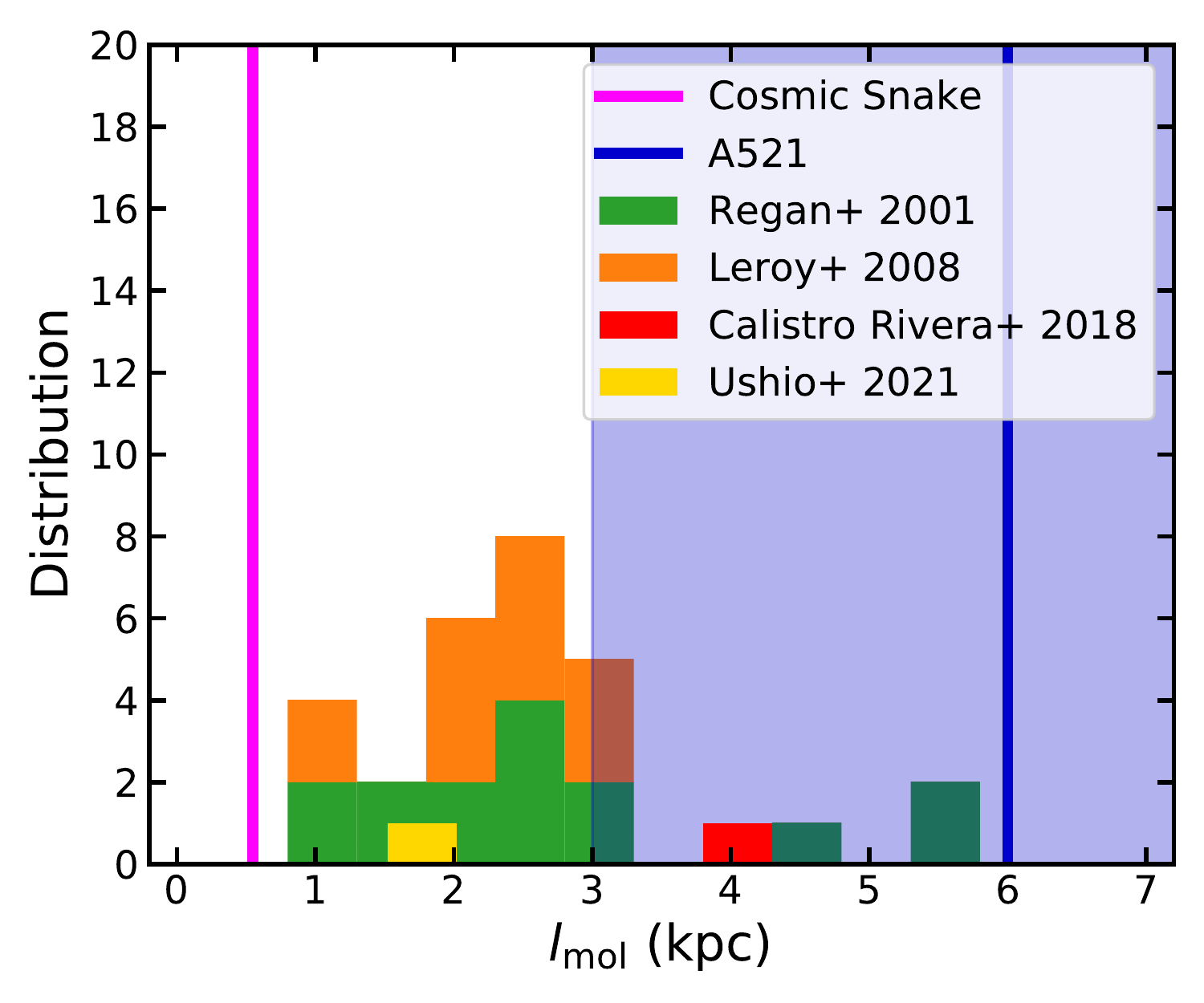}
      \end{subfigure}%
      \begin{subfigure}{0.333\textwidth}{}
        \includegraphics[width=\textwidth]{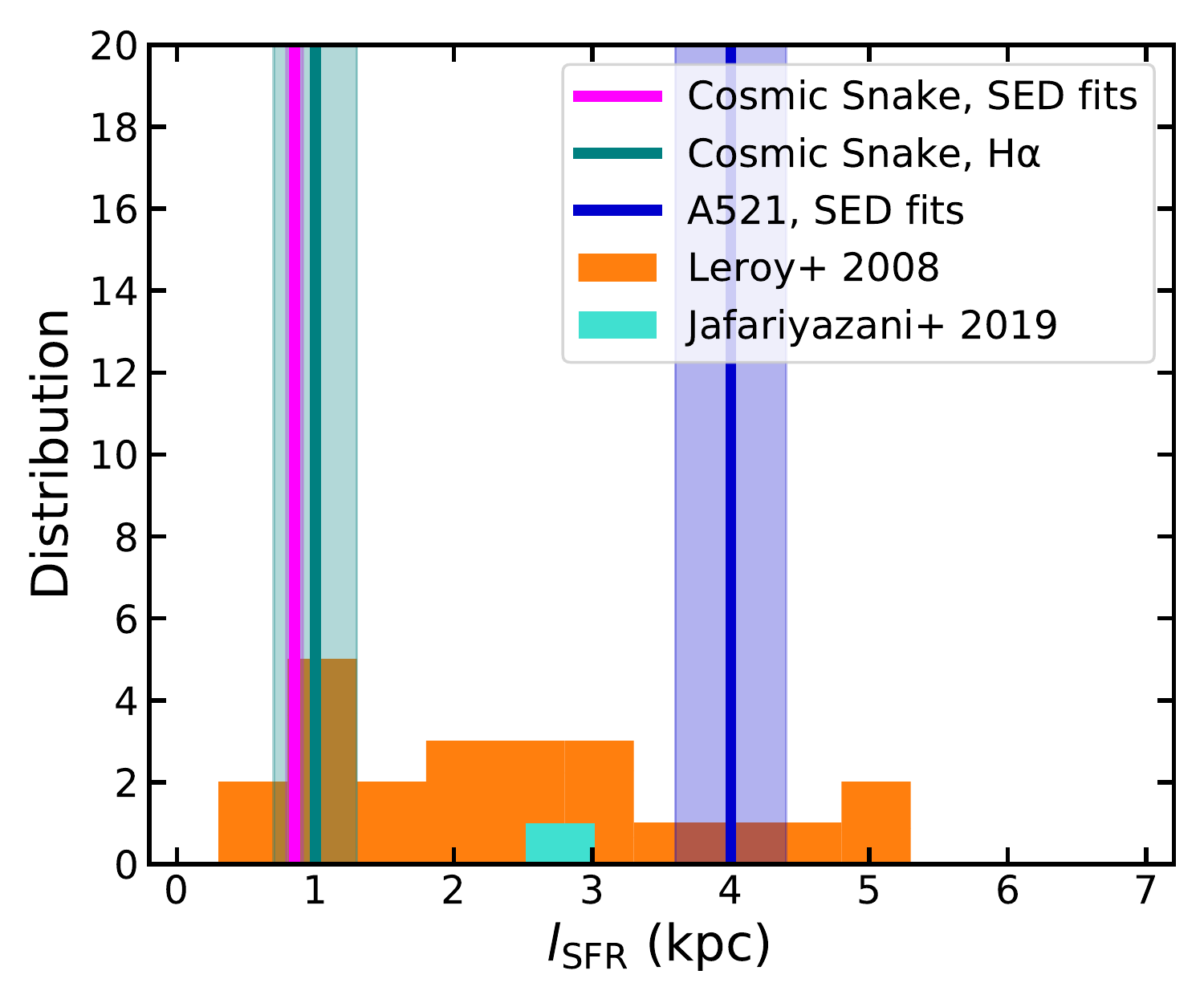}
      \end{subfigure}%
      \begin{subfigure}{0.333\textwidth}{}
        \includegraphics[width=\textwidth]{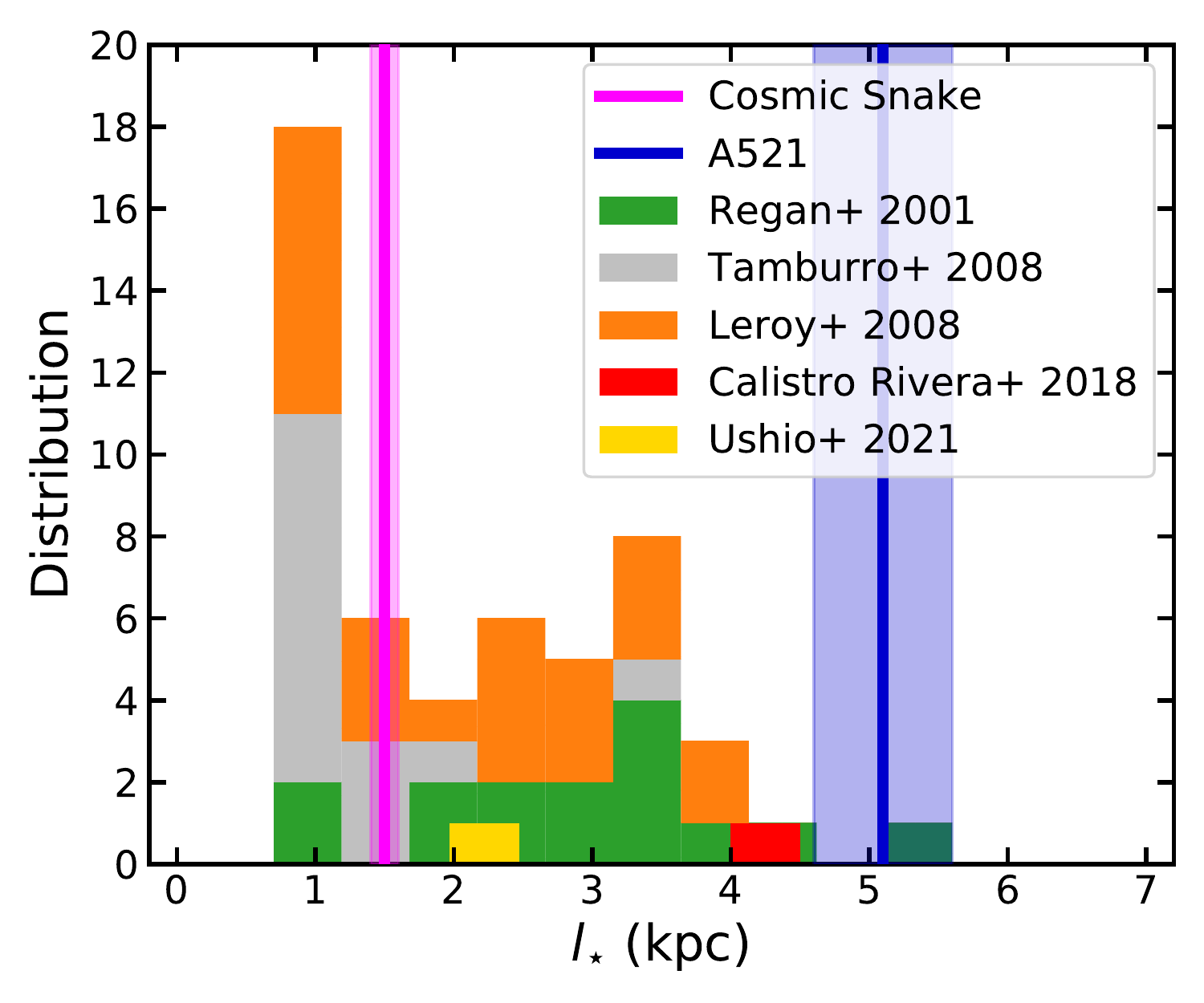}
      \end{subfigure}%
      
    \caption{Scale length comparison of $\Sigma M_{\mathrm{mol}}$ ($l_\mathrm{mol}$, left panel), $\Sigma \mathrm{SFR}$ ($l_\mathrm{SFR}$, middle panel), and $\Sigma M_\star$ ($l_\star$, right panel) of the Cosmic Snake and A521 with other studies from the literature: \citet{regan_bima_2001} in hatched green, \citet{tamburro_geometrically_2008} in hatched grey, and \citet{leroy_star_2008} in hatched orange for the $z=0$ galaxies, \citet{jafariyazani_spatially_2019} in turquoise for the $z=0.1-0.42$ galaxy stack, \citet{calistro_rivera_resolving_2018} in red for the $z=2-3$ sub-millimetre galaxy stack, and \citet{ushio_internal_2021} in yellow for the $z=1.45$ galaxy. Measurements for the Cosmic Snake and A521 are shown with magenta and blue vertical lines, respectively. Because we have two measurements of $\Sigma \mathrm{SFR}$, we plot the scale length for the Cosmic Snake derived from the $\mathrm{H\alpha}$ emission as a teal line. The magenta line is the scale length of the $\mathrm{\Sigma SFR}$ obtained from the SED fits. We use transparent rectangles around the vertical lines to visualise the uncertainty range.}\label{fig:histograms}
    \end{figure*}

In Fig. \ref{fig:histograms} we compare the scale lengths of $\Sigma \mathrm{SFR}$, $\Sigma M_\star$, and $\Sigma M_{\mathrm{mol}}$ of the Cosmic Snake and A521 with the scale lengths reported in the literature\footnote{Not all galaxies reported in the literature have measurements of the scale length of each physical quantity. Galaxies from \citet{regan_bima_2001}, \citet{ushio_internal_2021}, and \citet{calistro_rivera_resolving_2018} only have scale length measurements of $\Sigma M_{\mathrm{mol}}$ and $\Sigma M_{\star}$, \citet{tamburro_geometrically_2008} provided $\Sigma M_\star$ scale length measurements, and \citet{jafariyazani_spatially_2019} reported $\Sigma \mathrm{SFR}$ scale lengths.}. We considered 52 local galaxies, including 15 galaxies from \citet{regan_bima_2001}, 14 from \citet{tamburro_geometrically_2008}, and 23 from \citet{leroy_star_2008}. We also compared them with high-redshift galaxies with published radial profiles: a massive ($1.2\substack{+0.3\\-0.2}\times 10^{11}\,\si{M_\odot}$) MS galaxy at $z=1.45$ from \citet{ushio_internal_2021}, a stack of 32 galaxies at $z=0.1-0.42$ from \citet{jafariyazani_spatially_2019}, and a stack of 4 sub-millimetre galaxies at $z\sim 2-3$ from \citet{calistro_rivera_resolving_2018}. To trace the molecular gas, \citet{regan_bima_2001} used the CO(1--0) emission line, whereas \citet{leroy_star_2008} used the CO(2--1) emission in all but two galaxies (NGC 3627 and NGC 5194), where they used the CO(1--0) emission. \citet{ushio_internal_2021} used the CO(2--1) emission line, and \citet{calistro_rivera_resolving_2018} used the CO(3--2) emission line.

The distribution of the derived scale lengths in local galaxies is very wide. It ranges from $\sim \SI{0.5}{kpc}$ to $\sim \SI{5.5}{kpc}$ for all three quantities. Moreover, no specific trend emerges in the scale lengths as there is no significant peak in their distributions, with an exception at $l_\star \approx 1$. Compared to the average values from the sample of local galaxies, the Cosmic Snake has shorter scale lengths (with the exception of $l_\star$), while those of A521 are longer. Considering the broad distribution of scale lengths, the Cosmic Snake and A521 values still lie within the observed values for nearby galaxies. With $l_{\mathrm{SFR}}=4.0\pm \SI{0.4}{kpc}$, A521 indicates an important star-forming activity in its outer regions, most likely in the spiral arms. This is in line with the bump seen at $4-\SI{5}{kpc}$ in the $\Sigma M_{\mathrm{mol}}$ profile. These bumps are observed in the CO radial profiles of local galaxies as well (e.g. in \citet{leroy_star_2008}, in NGC 3198, NGC 3351, NGC 3627, and NGC 5194, or in \citet{regan_bima_2001}, in NGC 628) and are clearly associated with the spiral arms observed in the molecular gas maps of these galaxies. However, $l_{\mathrm{mol}}$ of these local galaxies is never as long as in A521, they are not longer than $\SI{3}{kpc}$, except in NGC 628 where $l_{\mathrm{mol}} = 5.8 \pm 0.2\,\si{kpc}$.

The Cosmic Snake and A521 both appear to be different from other high-redshift galaxies that we can use for a more direct comparison with contemporary galaxies: the scale lengths reported for the MS galaxy at $z=1.45$ by \citet{ushio_internal_2021} and for the stacks of galaxies at $z>0.1$ (\citealt{jafariyazani_spatially_2019,calistro_rivera_resolving_2018}) are shorter than in A521 and longer than in the Cosmic Snake for all three physical quantities. Altogether, this suggests that $l_{\mathrm{mol}}$, $l_{\mathrm{SFR}}$, and $l_{\star}$ of high-redshift ($z=0.1-3$) galaxies cover a similarly wide range of values as in local galaxies. There is no trend in a favoured scale length value, in line with our discussion in Section \ref{sec:morphology}. More measurements are necessary to support these preliminary results and to highlight possible morphological trends, such as spiral arms or starbursts (as in sub-millimetre galaxies), or any other trend tracing the evolutionary stage of the galaxy.

\begin{figure}
   \centering
   \includegraphics[width=0.4\textwidth]{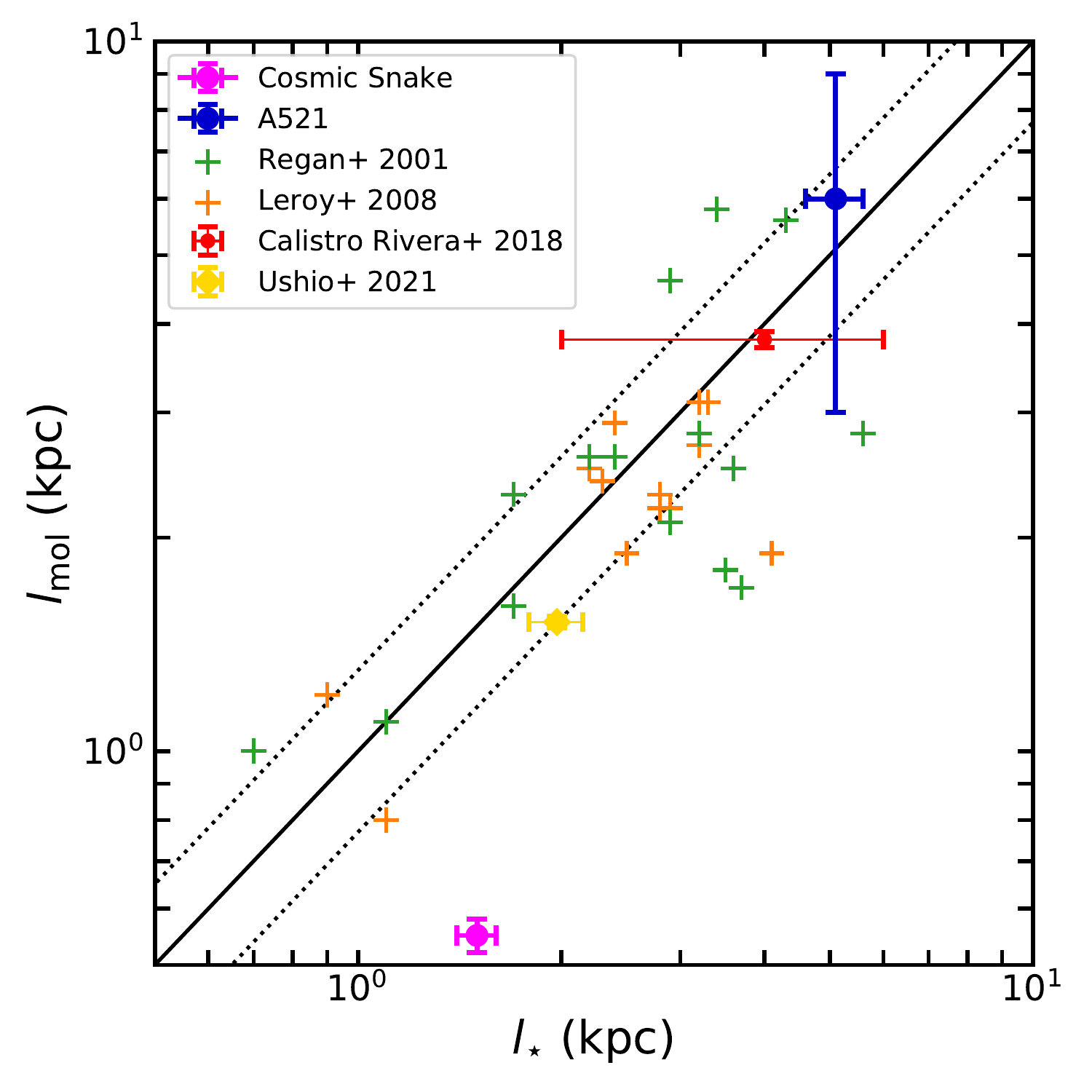}
      \caption{Scale lengths of molecular gas as a function of the stellar scale length for the Cosmic Snake and A521 compared with data from \citet{leroy_star_2008}, \citet{regan_bima_2001}, \citet{calistro_rivera_resolving_2018}, and \citet{ushio_internal_2021}. The continuous line shows equality, and the dotted lines show $\pm 30\%$.}
         \label{fig:covsstar}
   \end{figure}

As in \citet{leroy_star_2008} and \citet{regan_bima_2001}, we compared $l_\mathrm{mol}$ and $l_\star$ of our galaxies. The comparison with data from \citet{leroy_star_2008}, \citet{regan_bima_2001}, \citet{calistro_rivera_resolving_2018}, and \citet{ushio_internal_2021} is displayed in Fig. \ref{fig:covsstar}. For local galaxies, we expect a close to one-to-one relation (see \citealt{leroy_star_2008}). A521 satisfies the one-to-one relation, similarly to the stack from \cite{calistro_rivera_resolving_2018} and to the $z=1.45$ galaxy from \cite{ushio_internal_2021}. The Cosmic Snake, on the other hand, has a noticeably long $l_\star$ compared to $l_{\mathrm{mol}}$.

We can specifically compare the Cosmic Snake and A521 with local galaxies that have similar $\Sigma M_{\mathrm{mol}}$ and $\Sigma \mathrm{SFR}$ radial profiles (similar scale lengths), for example NGC 4736 and NGC 5194, respectively \citep{leroy_star_2008}. NGC 4736 has a quickly decreasing $\Sigma M_{\mathrm{mol}}$ profile like the Cosmic Snake, whereas in the $\Sigma M_{\mathrm{mol}}$ profile of NGC 5194, we note a bump at $\sim 4- \SI{5}{kpc}$ like in A521. The $\Sigma \mathrm{SFR}$ profiles of the two local galaxies are both relatively flat and bumpy as in the Cosmic Snake and A521. The bumps have been shown to be associated with bursty star formation in local galaxies. {The main difference between the Cosmic Snake and A521 and the selected $z=0$ counterparts (NGC 4736 and NGC 5194) is that the normalisation of the $\Sigma M_{\mathrm{mol}}$ profiles of the $z=1$ galaxies is offset towards high surface densities by $\sim 20$ times for the Cosmic Snake and $\sim 4-10$ times\footnote{Going from 4 times in the centre, up to more than 10 times at a galactocentric radius of $\SI{6}{kpc}$.} for A521, and this is similar for the $\Sigma \mathrm{SFR}$ radial profiles. Comparable $\Sigma M_{\mathrm{mol}}$ and $\Sigma \mathrm{SFR}$ offsets are observed between the radial profiles of any local galaxy \citep{regan_bima_2001,tamburro_geometrically_2008,leroy_star_2008} and the Cosmic Snake and A521}. Because a higher $M_{\mathrm{mol}}$ implies a higher SFR according to the KS law \citep{schmidt_rate_1959,schmidt_rate_1963,kennicutt_star_1989}, it is expected to find the $\Sigma \mathrm{SFR}$ radial profiles to be offset as well. This is indeed the case in both galaxies. Their $\Sigma M_{\mathrm{mol}}$ and $\Sigma \mathrm{SFR}$ radial profiles are offset by the same factors. This suggests a good agreement with the KS law derived for local galaxies: $\Sigma \mathrm{SFR} \propto (\Sigma M_{\mathrm{mol}})^n$, with $n=0.8-1.2$ (e.g. \citealt{bigiel_star_2008,bigiel_constant_2011,leroy_molecular_2013}). We defer the study of the resolved KS law in the Cosmic Snake and A521 to a future paper (Nagy et al. in prep.).

\section{Conclusions}
\label{sec:conclusion}

We have studied two strongly lensed galaxies at $z\sim 1$, the Cosmic Snake behind MACS J1206.2-0847, and A521 behind Abell 0521. Through gravitational lensing, we were able to access to sub-hundred parsec resolutions, which enabled us to undertake spatially resolved studies of these two galaxies.

We derived the radial profiles of $\Sigma M_\star$, $\Sigma \mathrm{SFR}$, and $\Sigma M_{\mathrm{mol}}$ for both galaxies, and for the Cosmic Snake alone, we also derived the radial profile of $\Sigma M_{\mathrm{dust}}$. $\Sigma M_\star$ and $\Sigma \mathrm{SFR}$ were obtained through SED fitting of the HST photometry of the galaxies. For the Cosmic Snake, we also used the $\mathrm{H\alpha}$ emission, mapped with SINFONI, as the tracer of SFR. $\Sigma M_{\mathrm{mol}}$ was derived using the CO(4--3) emission, observed with ALMA, as the tracer of $\mathrm{H}_2$. We performed MBB fitting on the FIR SED, using Herschel data and ALMA rest-frame $\SI{650}{\mu m}$ data. We then obtained the radial profile of $\Sigma M_{\mathrm{dust}}$ traced by the rest-frame $\SI{650}{\mu m}$ dust-continuum emission.

We fitted exponential functions on the radial profiles in order to derive the scale lengths of the physical quantities, and found significant differences between the two galaxies. The Cosmic Snake has much shorter scale lengths than A521, even though their integrated properties are similar.

We compared the Cosmic Snake and A521 with other galaxies, including 52 local galaxies from \citet{regan_bima_2001}, \citet{leroy_star_2008}, and \citet{tamburro_geometrically_2008}. In local galaxies, we found that the distribution of scale lengths is very wide and does not show a trend in a preferred scale length value. The Cosmic Snake lies at the short end of the local scale length distribution, whereas A521 is at the long end of the local scale length distribution. Moreover, we also included one individual MS galaxy at $z=1.45$ from \citet{ushio_internal_2021} and two stacks of galaxies at $z= 0.1-0.42$ from \citet{jafariyazani_spatially_2019}, and at $z\sim 2-3$ from \citet{calistro_rivera_resolving_2018} in the comparison. We conclude that high-redshift galaxies seem to have a similarly wide distribution of $\Sigma M_\star$, $\Sigma \mathrm{SFR}$, and $\Sigma M_{\mathrm{mol}}$ scale lengths, although more measurements are necessary to confirm this result.

For the SFR and the $M_{\mathrm{mol}}$ radial profiles and their scale lengths, we conclude that the large difference between the Cosmic Snake and A521 was not surprising. The radial profiles of these quantities might indeed vary strongly during the lifespan of a galaxy. However, the $M_\star$ distribution is more stable between galaxies, and in particular, $\mu_\star^{\mathrm{HLR}}$  mainly depends on the total $M_\star$. We therefore computed $\mu_\star^{\mathrm{HLR}}$ of the Cosmic Snake and A521 and found that both galaxies satisfy the $\mu_\star^{\mathrm{HLR}}-M_\star$ scaling relation that was empirically defined for local galaxies. Moreover, because this relation helps to determine the morphological type of galaxies, we were able to observe that the Cosmic Snake appears to be of later type than A521.

We found galaxies in the nearby universe with $\Sigma M_{\mathrm{mol}}$ and $\Sigma \mathrm{SFR}$ radial profiles very similar to the Cosmic Snake and A521, respectively. {The main difference is that the normalisation of the $\Sigma M_{\mathrm{mol}}$ and $\Sigma \mathrm{SFR}$ profiles of our $z\sim 1$ galaxies is offset by a factor of up to 20 towards higher values with respect to any $z=0$ galaxy}. This shows that $\Sigma M_{\mathrm{mol}}$ is significantly higher in the Cosmic Snake and A521, and the high $\Sigma \mathrm{SFR}$ is a direct consequence of this.  $\Sigma \mathrm{SFR}$ is indeed offset by the same factor as the $\Sigma M_{\mathrm{mol}}$, as expected from the KS law.

\begin{acknowledgements}
    This work was supported by the Swiss National Science Foundation.
    
    Based on observations made with the NASA/ESA Hubble Space Telescope, and obtained from the Data Archive at the Space Telescope Science Institute, which is operated by the Association of Universities for Research in Astronomy, Inc., under NASA contract NAS 5-26555. These observations are associated with program \#15435.
    
    This paper makes use of the following ALMA data: ADS/JAO.ALMA\#2013.1.01330.S, and ADS/JAO.ALMA\#2016.1.00643.S. ALMA is a partnership of ESO (representing its member states), NSF (USA) and NINS (Japan), together with NRC (Canada), MOST and ASIAA (Taiwan), and KASI (Republic of Korea), in cooperation with the Republic of Chile. The Joint ALMA Observatory is operated by ESO, AUI/NRAO and NAOJ.
    
    Also based on SINFONI observations made with the European Southern Observatory VLT telescope, Paranal, Chile, collected under the programme ID No. 087.A-0700.
    
    This work follows on from observations made with the Herschel Space Observatory, a European Space Agency Cornerstone Mission with significant participation by NASA. Support for this work was provided by NASA through an award issued by JPL/Caltech.
    
    We also used PdBI observations. PdBI is run by the Institut de Radioastronomie Millimétrique (IRAM, France), a partnership of the French CNRS, the German MPG and the Spanish IGN.
    
    M.M. acknowledges the support of the Swedish Research Council, Vetenskapsrådet (internationell postdok grant 2019-00502). 
\end{acknowledgements}

%
%

\begin{appendix} 

\section{Residuals of the HST images of the Cosmic Snake}
\label{app:residuals}

\begin{figure*}
    \centering
    \includegraphics[width=0.9\textwidth]{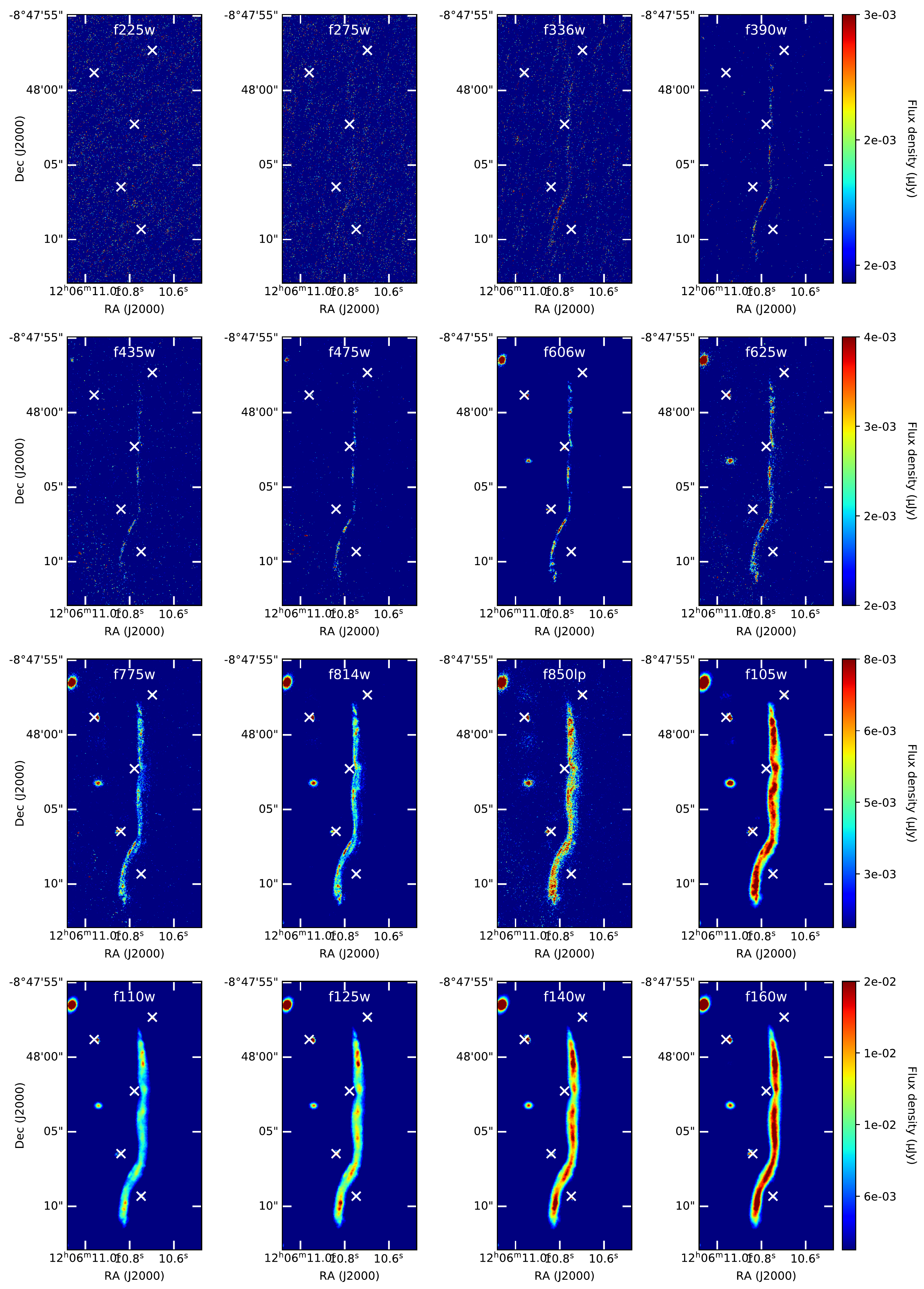}
    \caption{Cosmic Snake in 16 HST bands. Five cluster galaxies have been removed by subtracting fitted double 2D Sérsic functions. White crosses mark the location of the centre of the subtracted galaxies. The images are regrouped in lines of four panels that share the same flux range, labelled by colour bars on the right.
    }
    \label{fig:snake_residuals}

\end{figure*}

This appendix presents the modelling of the five closest and brightest cluster members to the Cosmic Snake arc. We also report the subtraction of their fluxes, which we performed to recover the intrinsic light of the Cosmic Snake. We fitted the light profile of each cluster member with a sum of two 2D Sérsic functions of the form

\begin{equation}
    I(x,y) = I(r) = I_e\exp\left\{-b_n\left[\left(\frac{r}{r_{e}}\right)^{(1/n)}-1\right]\right\}
,\end{equation}

\noindent where $I_e$ is the surface brightness at the effective radius $r_e$, $b_n$ is a parameter defined such that half of the total luminosity is emitted at $r_e$ and can be computed using $\Gamma(2n) = 2\gamma (2n,b_n)$, where $\Gamma(x)$ and $\gamma(s,x)$ are the gamma function and the lower incomplete gamma function, respectively. The fit minimisation was performed using the Levenberg-Marquardt algorithm and least-squares statistic. A mask was applied on the Cosmic Snake arc to exclude from the fit the regions at galactocentric radii below $\SI{4}{kpc}$ (chosen to be enough for the fit to converge), meaning that no pixel inside this region was considered for fitting. The position of the centre of a given galaxy was fixed on the position of its brightest pixel. The fit was performed recursively: the galaxies were subtracted one by one from the image, such that modifications made to the image by one subtraction were taken into account for the next fit. The residual images of the Cosmic Snake arc after subtracting the best fits are shown in Fig. \ref{fig:snake_residuals}.

To estimate the uncertainty ($\Delta S_{\mathrm{band}, \mathrm{ring}}$) on the flux density in the residual images in a given band and for a given elliptical ring, we computed the RMS of the residual map ($\sigma_{\mathrm{RMS}}$) inside $\mathrm{2\sigma}$ contours around the cluster galaxies. We discarded points closer than $\SI{10}{kpc}$ to the centres of the cluster galaxies because the accuracy of the fit near the peak does not affect the photometry of the Cosmic Snake. The error was additionally weighted by the normalised fitted model obtained through

\begin{equation}
w_{\mathrm{band}, \mathrm{ring}} = \frac{|S_{\mathrm{band}, \mathrm{ring}}^{\mathrm{raw}}-S_{\mathrm{band}, \mathrm{ring}}^{\mathrm{res}}|}{\mathrm{Max}_{\mathrm{ring}}(|S_{\mathrm{band}, \mathrm{ring}}^{\mathrm{raw}}-S_{\mathrm{band}, \mathrm{ring}}^{\mathrm{res}}|)}
,\end{equation}

\noindent with the labels `raw' and `res' referring to flux densities of the raw image and the residual map\footnote{The function $\mathrm{Max}_{\mathrm{ring}}(x_{\mathrm{band}, \mathrm{ring}})$ gives the maximum value of $x_{\mathrm{band}, \mathrm{ring}}$ for a fixed band.}. Including $w_{\mathrm{band}, \mathrm{ring}}$ ensures that the error is larger in regions that are strongly affected by the fit. The final error is thus given by $\Delta S_{\mathrm{band}, \mathrm{ring}} = \sigma_{\mathrm{RMS}} \cdot w_{\mathrm{band}, \si{ring}}$.

\section{Photometry}

We show in Table \ref{table:photometrysnake} and Table \ref{table:photometryA521} the photometries of the Cosmic Snake and A521, respectively. The columns correspond to the HST bands, and the rows correspond to different radii of the rings. The rings have widths of $\SI{0.35}{kpc}$ and $\SI{0.4}{kpc}$ for the Cosmic Snake and A521, respectively. The value shown in the first column (r (kpc)) is the outer radius of the ring. The uncertainty on a given surface brightness is displayed as a number in parenthesis, corresponding to the error on the last digit (e.g. $0.014(3) = 0.014 \pm 0.003$).

\begin{table*}
    \caption{Photometry of the Cosmic Snake in 16 HST bands. The values are in units of $\si{\mu Jy.kpc^{-2}}$}             
    \label{table:photometrysnake}      
    \centering          
    \begin{tabular}{cllllllll}
    \hline\hline
    r (kpc) & F225W & F275W & F336W & F390W & F435W & F475W & F606W & F625W \\
    \hline
    0.35 & 0.003(2) & 0.004(2) & 0.003(1) & 0.0033(7) & 0.0029(8) & 0.0033(7) & 0.009(1) & 0.010(1) \\
    0.70 & 0.003(1) & 0.002(1) & 0.0043(8) & 0.0034(6) & 0.0033(6) & 0.0048(7) & 0.010(2) & 0.013(1) \\
    1.05 & 0.002(1) & 0.0025(8) & 0.0041(8) & 0.0047(5) & 0.0039(5) & 0.0060(7) & 0.012(2) & 0.016(2) \\
    1.40 & 0.0026(9) & 0.0029(7) & 0.0045(9) & 0.0057(5) & 0.0056(4) & 0.0068(7) & 0.014(2) & 0.019(2) \\
    1.75 & 0.0022(9) & 0.0039(8) & 0.006(1) & 0.0079(5) & 0.0068(5) & 0.0095(7) & 0.017(2) & 0.020(2) \\
    2.10 & 0.0025(8) & 0.0046(8) & 0.007(1) & 0.0090(5) & 0.0092(5) & 0.0117(7) & 0.018(2) & 0.021(2) \\
    2.45 & 0.0031(7) & 0.0065(8) & 0.008(1) & 0.0105(5) & 0.0105(5) & 0.0126(7) & 0.018(2) & 0.022(3) \\
    2.80 & 0.0030(7) & 0.0057(8) & 0.009(1) & 0.0101(5) & 0.0095(5) & 0.0116(7) & 0.016(2) & 0.020(2) \\
    3.15 & 0.0031(8) & 0.0048(9) & 0.006(1) & 0.0075(5) & 0.0078(4) & 0.0090(7) & 0.013(2) & 0.016(2) \\
    3.50 & 0.0029(7) & 0.0035(8) & 0.0038(8) & 0.0045(5) & 0.0043(4) & 0.0058(7) & 0.008(1) & 0.010(1) \\
    3.85 & 0.0021(7) & 0.0027(8) & 0.0025(7) & 0.0028(5) & 0.0022(4) & 0.0033(5) & 0.0050(8) & 0.006(1) \\
    4.20 & 0.0014(7) & 0.0019(7) & 0.0024(8) & 0.0017(4) & 0.0018(4) & 0.0021(5) & 0.0031(6) & 0.0046(9) \\
    4.55 & 0.0015(7) & 0.0021(6) & 0.008(1) & 0.0013(3) & 0.0017(4) & 0.0015(3) & 0.0034(8) & 0.0027(8) \\
    4.90 & 0.0020(7) & 0.0026(7) & 0.0018(7) & 0.0011(3) & 0.0013(4) & 0.0015(3) & 0.0022(6) & 0.0024(8) \\
    5.25 & 0.0023(8) & 0.0018(6) & 0.0020(7) & 0.0012(3) & 0.0016(4) & 0.0010(3) & 0.0016(4) & 0.0022(7) \\
    5.60 & 0.0020(8) & 0.0021(8) & 0.0019(6) & 0.0011(3) & 0.0014(4) & 0.0010(3) & 0.0009(3) & 0.0022(6) \\
    5.95 & 0.0022(7) & 0.0015(8) & 0.0016(6) & 0.0012(4) & 0.0011(4) & 0.0010(3) & 0.0008(4) & 0.0018(6) \\
    6.30 & 0.0033(9) & 0.0014(8) & 0.0011(5) & 0.0009(3) & 0.0011(4) & 0.0012(4) & 0.0008(4) & 0.0015(6) \\
    6.65 & 0.0023(7) & 0.0025(7) & 0.0015(7) & 0.0010(3) & 0.0014(4) & 0.0009(3) & 0.0008(3) & 0.0018(6) \\
    7.00 & 0.0012(8) & 0.0017(6) & 0.0012(5) & 0.0007(3) & 0.0006(5) & 0.0010(3) & 0.0008(4) & 0.0022(6) \\
    \hline\hline
    r (kpc) & F775W & F814W & F850LP & F105W & F110W & F125W & F140W & F160W \\
    \hline
    0.35 & 0.028(3) & 0.040(5) & 0.066(9) & 0.105(9) & 0.15(2) & 0.18(1) & 0.23(2) & 0.28(3) \\
    0.70 & 0.034(4) & 0.043(5) & 0.07(1) & 0.105(9) & 0.14(1) & 0.18(1) & 0.22(2) & 0.26(2) \\
    1.05 & 0.038(5) & 0.048(6) & 0.070(9) & 0.104(9) & 0.14(2) & 0.17(1) & 0.20(2) & 0.23(3) \\
    1.40 & 0.040(4) & 0.050(6) & 0.074(9) & 0.102(9) & 0.13(1) & 0.15(1) & 0.18(2) & 0.21(2) \\
    1.75 & 0.042(6) & 0.051(7) & 0.072(9) & 0.094(9) & 0.12(1) & 0.14(1) & 0.16(2) & 0.18(2) \\
    2.10 & 0.041(5) & 0.048(6) & 0.066(9) & 0.084(9) & 0.10(1) & 0.12(1) & 0.14(2) & 0.15(2) \\
    2.45 & 0.040(5) & 0.047(6) & 0.061(9) & 0.073(9) & 0.09(1) & 0.10(1) & 0.12(2) & 0.12(1) \\
    2.80 & 0.035(4) & 0.041(6) & 0.054(7) & 0.061(8) & 0.074(9) & 0.08(1) & 0.09(1) & 0.10(1) \\
    3.15 & 0.028(3) & 0.032(4) & 0.043(6) & 0.050(8) & 0.059(7) & 0.06(1) & 0.08(1) & 0.08(1) \\
    3.50 & 0.018(2) & 0.021(3) & 0.030(3) & 0.036(7) & 0.046(7) & 0.048(7) & 0.058(9) & 0.060(8) \\
    3.85 & 0.012(2) & 0.015(3) & 0.022(3) & 0.027(4) & 0.034(4) & 0.034(3) & 0.040(5) & 0.044(4) \\
    4.20 & 0.009(1) & 0.010(1) & 0.015(1) & 0.018(2) & 0.025(4) & 0.025(3) & 0.026(5) & 0.030(4) \\
    4.55 & 0.007(1) & 0.008(1) & 0.011(1) & 0.013(2) & 0.015(3) & 0.019(3) & 0.022(5) & 0.022(3) \\
    4.90 & 0.008(1) & 0.006(1) & 0.008(1) & 0.011(2) & 0.009(2) & 0.012(2) & 0.018(3) & 0.018(4) \\
    5.25 & 0.0040(9) & 0.0049(9) & 0.007(1) & 0.009(1) & 0.008(2) & 0.010(1) & 0.014(3) & 0.015(3) \\
    5.60 & 0.0041(7) & 0.0045(9) & 0.0067(9) & 0.008(1) & 0.008(2) & 0.009(1) & 0.013(2) & 0.013(2) \\
    5.95 & 0.0029(7) & 0.0032(7) & 0.0042(9) & 0.007(1) & 0.0070(9) & 0.008(1) & 0.010(2) & 0.011(1) \\
    6.30 & 0.0034(8) & 0.0029(5) & 0.0043(9) & 0.0058(6) & 0.0055(5) & 0.007(1) & 0.0084(4) & 0.010(2) \\
    6.65 & 0.0034(7) & 0.0024(4) & 0.0035(9) & 0.0053(7) & 0.0044(6) & 0.005(1) & 0.0079(4) & 0.009(2) \\
    7.00 & 0.0032(8) & 0.0024(4) & 0.0045(9) & 0.0056(9) & 0.0034(3) & 0.006(1) & 0.0086(4) & 0.009(2) \\
    \hline               
    \end{tabular}
\end{table*}

\begin{table*}
\caption{Photometry of A521 in 4 HST bands. The values are in units of $\si{\mu Jy.kpc^{-2}}$}             
\label{table:photometryA521}      
\centering          
\begin{tabular}{cllll}

\hline\hline
r (kpc) & F390W & F606W & F105W & F160W \\
\hline
0.4 & 0.011(1) & 0.034(3) & 0.185(3) & 0.351(4) \\
0.8 & 0.014(2) & 0.037(3) & 0.179(3) & 0.338(5) \\
1.2 & 0.017(2) & 0.035(2) & 0.161(3) & 0.299(4) \\
1.6 & 0.017(2) & 0.048(5) & 0.161(4) & 0.282(4) \\
2.0 & 0.018(3) & 0.046(6) & 0.148(4) & 0.268(5) \\
2.4 & 0.029(5) & 0.055(8) & 0.149(5) & 0.252(4) \\
2.8 & 0.024(5) & 0.044(7) & 0.132(3) & 0.229(3) \\
3.2 & 0.016(2) & 0.032(3) & 0.126(3) & 0.221(3) \\
3.6 & 0.012(1) & 0.027(2) & 0.113(2) & 0.205(2) \\
4.0 & 0.0099(8) & 0.025(1) & 0.106(2) & 0.194(2) \\
4.4 & 0.0097(5) & 0.026(1) & 0.104(2) & 0.188(3) \\
4.8 & 0.0113(6) & 0.023(1) & 0.096(2) & 0.167(3) \\
5.2 & 0.0109(5) & 0.0220(9) & 0.088(2) & 0.155(2) \\
5.6 & 0.0087(5) & 0.019(1) & 0.079(1) & 0.141(2) \\
6.0 & 0.0063(3) & 0.0176(9) & 0.069(1) & 0.124(2) \\
6.4 & 0.0054(4) & 0.016(1) & 0.064(1) & 0.115(2) \\
6.8 & 0.0060(4) & 0.0127(9) & 0.055(1) & 0.099(2) \\
7.2 & 0.0053(4) & 0.0115(8) & 0.050(1) & 0.089(2) \\
7.6 & 0.0044(4) & 0.012(1) & 0.048(1) & 0.085(2) \\
8.0 & 0.0044(5) & 0.0100(9) & 0.042(1) & 0.074(2) \\
\hline

\end{tabular}
\end{table*}

\end{appendix}

\end{document}